\documentstyle[12pt,amsfonts,eucal]{article}

\newcommand{\bra}{\langle}
\newcommand{\ket}{\rangle}
\newcommand{\CC}{{\mathbb C}}

\newcommand{\dd}{\partial}
\newcommand{\DD}{{\cal D}}

\newcommand{\GG}{{\cal G}}
\newcommand{\Gr}{{\rm Gr}\,}
\newcommand{\HP}{{\bf H}}
\newcommand{\HH}{{\cal H}}
\newcommand{\jj}{$j=1,\,\ldots,\,n$}

\newcommand{\Kr}{{\rm Ker}}

\newcommand{\N}{{\cal N}}
\newcommand{\Q}{{\cal Q}}
\newcommand{\RR}{{\mathbb R}}
\newcommand{\Rs}{{\mathbb R}_+}
\newcommand{\SS}{{\mathbb S}}
\newcommand{\TT}{{\mathbb T}}
\newcommand{\Tr}{{\rm Tr}}
\newcommand{\tQ}{{\tilde Q}}

\newcommand{\zep}{z_{\epsilon}}
\newcommand{\ZZ}{{\mathbb Z}}

\newcommand{\va}{{\vec a}}
\newcommand{\vb}{{\vec b}}
\newcommand{\ee}{{\bf e}}
\newcommand{\ba}{{\bf a}}
\newcommand{\bb}{{\bf b}}
\newcommand{\Ker}{{\rm Ker}}
\newcommand{\Ran}{{\rm Ran}}

\newcommand{\halm}{{\vrule height7pt width4pt depth0pt}}

\renewcommand{\Re}{{\rm Re}}
\renewcommand{\epsilon}{{\varepsilon}}
\renewcommand{\phi}{{\varphi}}
\renewcommand{\hat}{\widehat}
\renewcommand{\tilde}{\widetilde}
\renewcommand{\det}{{\rm det}}

\title
{\normalsize \bf
\vskip 2truecm
Scattering on compact manifolds with infinitely thin horns}
\author
{\normalsize
Jochen~BR\"UNING and  Vladimir~GEYLER}

\date{}

\begin{document}

\maketitle
\thispagestyle{empty}

\baselineskip=12pt
\begin{quote}

\end{quote}

\vskip 1truecm
\baselineskip=15pt

\noindent{\bf 0.~Introduction}

\bigskip

\noindent In the paper \cite{Fad} L.~D.~Faddeev initiated the investigation
of the quantum mechanical scattering on manifolds of constant negative
curvature with cusps (sometimes also called "horns" \cite{Gut1}); further
developments of this theory are presented e.g. in \cite {EGM}, \cite{FP},
\cite{LP}, \cite{Ven}. It is interesting to note that an explicit
expression for the reflection coefficient in the case of one horn was
obtained earlier by R.~Godement \cite{God}. Note also that M.~G.~Gutzwiller
has revealed a relation between the scattering theory on manifolds with
horns and the description of chaotic behavior of quantum systems
\cite{Gut1}, \cite{Gut2}, \cite{Gut3}.

If we imagine the width of the horns tending to zero, then we obtain
a so-called hedgehog-shaped topological space (or ``horned manifold''). 
Strictly speaking, we consider the limit of a family
of horned spaces in the sense of the Hausdorff--Gromov distance \cite{Gro}.
The simplest specimen of such a manifold is the Euclidean plane with an
attached half-line. The quantum mechanical scattering in this system has
been investigated for the first time by P.~Exner and P.~\v Seba \cite{ES1};
in \cite{ES2} these authors consider a compact plane domain
with a half-line glued to it. A series of significant physical
applications of the corresponding results as well as an intensive
bibliography related to the subject in question may be found in \cite{ETV};
we may add that the considered problem is also connected with the scattering
on graphs \cite{ACDMT}, \cite{ES3}, \cite{GPav}, \cite{KS}, \cite{Nov}.
An explicit expression for the
transmission coefficient in the case of two half-lines (``wires'') 
attached to a
compact Riemannian manifold of dimension two or three with some special
boundary conditions at the points of gluing has been obtained by
A.~Kiselev \cite{Kis}.
A general method of solving the transmission problem through an arbitrary
quantum device was
proposed in \cite{GP}, this method is based on an approach to modeling of
quantum systems developed by B.~S.~Pavlov \cite{Pav}; some of its
applications are given e.g. in \cite{BMPY}, \cite{GDM}, \cite{GPP}, 
\cite{PPGP}, \cite {PGP}.
Many-terminal problems for a bounded domain in $\RR^d$ ($d=2$ or $3$)
with wires attached to the boundary of the domain are considered recently 
in \cite{MP}, \cite{MPPRY}. 

In this paper we consider the quantum mechanical scattering in a
hedgehog-shaped space which is constructed by gluing a finite number of
half-lines to distinct points of a compact Riemannian manifold of dimension
less than four. The Hamiltonian of a quantum particle in such a system
coincides with a Schr\"odinger operator on the punctured manifold (the
points of gluing are removed) and with the free Schr\"odinger operator on
each half-line. At the gluing points, some boundary conditions are imposed.
In particular, the Schr\"odinger operator in a magnetic field is included
in our scheme. The approach we use is based on the Krein resolvent formula
from operator extension theory \cite{Kre}, therefore in Sec.~1 we give
a very brief sketch of results needed from this theory. Sec.~2 is devoted
to the construction of  Schr\"odinger operators on the hedgehog-shaped space;
we use the theory of boundary value spaces \cite{GG} to describe all
possible kinds of boundary conditions defining the Schr\"odinger operators.
We distinguish among them operators of "Dirichlet" and of "Neumann"
type. It is worth  noting that the results of Sec.~2 are valid
for all Riemannian manifolds of dimension less than four, not only for the
compact ones. In principle, the definition of the Schr\"odinger operator on
a hedgehog-shaped space may be given in the framework of pseudo-differential
operator theory on such a space \cite{Sch}, but our approach is more
convenient for investigating the scattering parameters and connected with
the approach to spectral problems for point perturbations on Riemannian
manifolds \cite{BG1}, \cite{BG2}.

The main results of the paper are contained in Sections~3 and 4. Here we get
a complete description of the spectral structure of Schr\"odinger operators
on hedgehog-shaped spaces (Theorem~4),
the proof of existence and
uniqueness of scattering states (Theorem~5), and the proof of the
unitary nature of the scattering matrix (Theorem~6). An explicit form
for the scattering matrix is given in the cases of arbitrary
Schr\"odinger operator on the
hedgehog shaped space (Formula (\ref{n3.40d})). In the particular case of the boundary
conditions of Neumann types our formulas contain the result of \cite{Kis} as
a very special case. Theorem~7 from Sec.~4 shows that the positive part of
the spectrum of the initial Schr\"odinger operator on the compact manifold
as well as the spectrum of a point perturbation of such an operator may be
recovered from the scattering amplitude for one attached half-line (so,
an "infinitely thin horn" may be considered as a kind of "quantum
stethoscope"). Moreover, the positive part of the spectrum of the initial
Schr\"odinger operator is fully determined by the conductance properties
of an "electronic device" consisting of the initial manifold and two
"wires" attached to it (Propositions~8 and 9).
We can choose the boundary conditions in such a way, that
in the limiting case
when wires are attached at the same point, the scattering matrix coincides
with that for the $\delta'$-interaction on the line. This fact is
related to a conjecture from \cite{AEL}: the scattering on the
"$\delta'$-potential" may be realized geometrically.
Finally, in Sec.~5 we give a series of
examples in which the Krein $\Q$-function entering our expressions for
the scattering matrix may be obtained in explicit form.
Note that some applications of the results of this paper to the 
conductance of the quantum sphere were considered recently in \cite{BGMP}.
Some aspects of the geometric scattering on non-compact Riemannian
manifolds within the framework of the approach presented here
are discussed in \cite{BG3}.

\vskip 5ex

\noindent{\bf 1.~Preliminaries}

\bigskip

\noindent Here we rephrase some results of operator extension theory
using the language of boundary value spaces and linear
symplectic  geometry (see, e.g. \cite{DM}, \cite{GG}, \cite{Har1},
\cite{Har2}, \cite{Koc1}, \cite{Pav} for details).

Let $V$ be a complex vector space with a skew-Hermitian sesquilinear
form $[x|y]$. The orthogonality with respect to this form will be denoted by
$[\bot]$: $x[\bot]y$ means that $[x|y]=0$;
the orthogonal complement of a set $X\subset V$ is denoted as $X^{[\bot]}$.
A subspace $\Lambda\subset V$ is called {\it isotropic}
(respectively, {\it Lagrangian}) if $\Lambda \subset \Lambda^{[\bot]}$
(respectively, $\Lambda = \Lambda^{[\bot]}$). If $\HH$ is a Hilbert space
with the scalar product $\bra x|y\ket$\footnote{Throughout this paper,
we assume that the scalar product is linear with respect to the {\it second}
argument}, then the Hilbert space $\HH\oplus\HH$ is endowed with the
{\it standard} skew-Hermitian form
%
%
%
$$
[x|y]=\bra x_1|y_2\ket-\bra x_2|y_1\ket\,,
$$
%
%
i.e. $[x|y]=\bra x|Jy\ket$, where $J:\HH\oplus\HH\to \HH\oplus\HH$ is a
unitary operator of the form $J(x_1,x_2)=(x_2,-x_1)$. It is clear that
$[x|y]$ is a continuous sesquilinear form on the Hilbert space
$\HH\oplus\HH$, hence, every Lagrangian subspace in $\HH\oplus\HH$ is closed.
Moreover, for every subset $X\subset \HH\oplus\HH$ we have
$X^{[\bot]}=(JX)^{\bot}=J(X^{\bot})$, where $X^{\bot}$ is the orthogonal
complement with respect to the standard scalar product $\bra x|y\ket$
in $\HH\oplus\HH$: $\bra x|y\ket=\bra x_1|y_1\ket+\bra x_2|y_2\ket$.
Therefore, a subspace $\Lambda\subset \HH\oplus\HH$ is isotropic
(respectively, Lagrangian) iff $J\Lambda\subset\Lambda^{\bot}$
(respectively, $J\Lambda=\Lambda^{\bot}$).

For every skew-Hermitian sesquilinear form $[x|y]$ the form $i\,[x|y]$
is Hermitian; therefore the geometry of a skew-Hermitian sesquilinear
form does not differ from that of a Hermitian form. Nevertheless, the
symplectic language is very useful in operator extension theory. For example,
let $A:\DD(A)\to \HH$ be a densely defined linear operator in $\HH$ with
the graph $\Gr(A)$, $\Gr(A)\subset \HH\oplus\HH$. Then it is easy to check
the following statements:

\medskip

\noindent (1) {\it $A$ is symmetric if
and only if $\Gr(A)$ is an isotropic subspace of $\HH\oplus\HH$}.

\noindent (2) {\it $A$ is self-adjoint if and only if
$\Gr(A)$ is a Lagrangian subspace of $\HH\oplus\HH$}.

\medskip

\noindent{\bf Remark 1}.
It is clear that every Lagrangian subspace is a maximal isotropic subspace,
the converse is not true even in the one-dimensional case.
On the other hand, if $V$ is a
finite-dimensional complex space having at least one Lagrangian subspace,
then according to the Witt theorem \cite{Bou}, every maximal isotropic
subspace is Lagrangian. Therefore, in the finite-dimensional space
$V=\HH\oplus\HH$ every maximal isotropic subspace is Lagrangian.
On the contrary, let $\HH$ be an infinite-dimensional Hilbert space, and
let $A$ be a maximal symmetric operator in $\HH$ which is not self-adjoint.
Then ${\rm Gr}(A)$ is a maximal isotropic subspace of $\HH\oplus\HH$
which is not Lagrangian.

\medskip

A linear mapping $u: V_1\to V_2$ of complex vector spaces $V_1$, $V_2$ with
skew-Hermitian forms $[\,\cdot\,|\,\cdot\,]_1$, $[\,\cdot\,|\,\cdot\,]_2$,
respectively, is called {\it skew-unitary} if $[u(x)|u(y)]_2=[x|y\,]_1$
$\forall\, x,y\in V_1$. Now let $S$ be a symmetric operator in $\HH$;
in the graph $\Gr(S^*)$  of $S^*$ we shall consider the skew-Hermitian
form induced by the standard form from  $\HH\oplus\HH$. A pair
$(\GG,\,\Gamma)$, where $\GG$ is a Hilbert space and $\Gamma$ is a surjective
skew-unitary mapping from $\Gr(S^*)$ onto $\GG\oplus\GG$ is called
{\it a boundary value space} for $S$. It is known that a boundary value
space for $S$ exists if and only if the deficiency indices $n_+(S)$ and
$n_-(S)$ for $S$ coincide, i.e. if and only if $S$
has a self-adjoint extension. If this is the case and $(\GG,\,\Gamma)$
is a boundary value space for $S$, then ${\rm dim}\,\GG=n_+(S) (=n_-(S))$
and $\Gamma$ is a continuous operator with respect to the standard
Hilbert space topologies in $\Gr(S^*)$ and $\GG\oplus\GG$. Let
$\widehat \Gamma:\DD(S^*)\to \GG\oplus\GG$ be the composition of the
canonical bijection $\DD(S^*)\to \Gr(S^*)$ ($x\mapsto(x,S^*x)$) and $\Gamma$;
it is clear that $\widehat\Gamma$ is surjective. Moreover, if $\DD(S^*)$
is endowed with the graph scalar product
$\bra x|y\ket_S=\bra x|y\ket+\bra S^*x|S^*y\ket$, then  $\widehat \Gamma$
is continuous. Denote by $P_1$ and $P_2$ the canonical projections of
$\GG\oplus\GG$ onto $\GG\oplus\{0\}$ and $\{0\}\oplus\GG$,
respectively,  and by $\Gamma^{(1)}$, $\Gamma^{(2)}$ the operators
$P_1\widehat\Gamma$ and $P_2\widehat\Gamma$, respectively. Then for all
$x,y\in\DD(S^*)$ the following relation takes place:
\begin{equation}
                                          \label{n1.2}
\bra x|S^*y\ket-\bra S^*x|y\ket=
\bra\Gamma^{(1)}x|\Gamma^{(2)}y\ket-\bra\Gamma^{(2)}x|\Gamma^{(1)}y\ket\,.
\end{equation}

\noindent Conversely, a triple $(\GG,\,\Gamma^{(1)},\,\Gamma^{(2)})$,
where $\Gamma$ is a Hilbert space and
$\Gamma^{(j)}:\DD(S^*)\to \GG$ ($j=1,2$) are linear operators,
uniquely defines a boundary value space, if the mapping
$\DD(S^*)\ni x \mapsto (\Gamma^{(1)}x,\,\Gamma^{(2)}x)\in\GG\oplus\GG$
is surjective and the condition (\ref{n1.2}) holds. Indeed, it is
sufficient to define $\Gamma$
by the rule $\Gamma(x,S^*x)=(\Gamma^{(1)}x,\,\Gamma^{(2)}x)$.
The triple $(\GG,\,\Gamma^{(1)},\,\Gamma^{(2)})$
is also called a boundary value space for $S$.

The following theorem describes all self-adjoint extensions of $S$ with
help of the boundary value space.

\medskip

{\bf Theorem A}. {\it Let $S$  be a symmetric operator in a Hilbert
space $\HH$ with coinciding deficiency indices, and let $(\GG,\,\Gamma)$
be a boundary value space for $S$. Then for every Lagrangian subspace
$\Lambda\subset \GG\oplus\GG$ the set $\Gamma^{-1}(\Lambda)$ is the
graph of a self-adjoint operator $H^{\Lambda}$ that is a self-adjoint
extension of $S$. Moreover, the correspondence $\Lambda\mapsto H^{\Lambda}$
is a bijection between all Lagrangian subspaces of $\GG\oplus\GG$ and
all self-adjoint extensions of $S$}. \halm

\medskip

In other words, the self-adjoint extension  $H^{\Lambda}$ is defined by
the boundary condition
\begin{equation}
                                    \label{n1.3}
(\Gamma^{(1)}x,\Gamma^{(2)}x)\in\Lambda\,.
\end{equation}
More precisely, the domain of $H^{\Lambda}$ is the subspace of $\DD(S^*)$
given by $\DD(H^{\Lambda})=\{x\in \DD(S^*):\,(\Gamma^{(1)}x,
\Gamma^{(2)}x)\in\Lambda\}$, and $H^{\Lambda}$ is the restriction of $S^*$
to $\DD(H^{\Lambda})$. Condition (\ref{n1.3}) can be written in a more
convenient "operator" form. Namely, for every Lagrangian subspace
$\Lambda\subset\GG\oplus\GG$ there exists a uniquely defined unitary operator
$U_\Lambda$ acting in $\GG$ such that the relations
$(x_1,x_2)\in\Lambda$ and $i(I+U_\Lambda)x_1=(I-U_\Lambda)x_2$
are equivalent; $U_\Lambda$ is called the {\it Cayley transform} for $\Lambda$.
(If $\Lambda$ is the graph of a
self-adjoint operator $L$ in $\GG$, then $U_\Lambda$ is just the Cayley
transform for $L$). Moreover, the correspondence $\Lambda\mapsto U_\Lambda$
is a bijection between the sets of all Lagrangian subspaces of
$\GG\oplus\GG$ and all unitary operators in $\GG$. Using the notations
above we can rewrite  condition (\ref{n1.3}) in the desired operator form:
%
%
 %
$$                                 \label{n1.4}
(I-U_\Lambda)\Gamma^{(2)}x=i(I+U_\Lambda)\Gamma^{(1)}x\,.
$$
%
%
It is clear that a given Lagrangian subspace $\Lambda\subset\GG\oplus\GG$
has different equations of the form
$Lx_1=Mx_2$, where $L$ and $M$ are bounded linear operators in $\GG$.
Hence, a given boundary condition $(\Gamma^{(1)}x,\,\Gamma^{(2)}x)\in\Lambda$
may be represented in the operator form
\begin{equation}
                                  \label{n1.4a}
L\Gamma^{(1)}x=M\Gamma^{(2)}x
\end{equation}
in many ways. Denote by $A(L,M)$ the bounded operator from
$\GG\oplus\GG$ to $\GG$ taking $x=(x_1,x_2)\in\GG\oplus\GG$ to
$Lx_1-Mx_2\in\GG$.

\medskip

{\bf Proposition B}. {\it Let $L,M:\,\GG\to\GG$ be bounded linear operators.
The subspace $\Lambda$ of $\GG\oplus\GG$ determined by the equation
$Lx_1=Mx_2$ is Lagrangian if and only if the following conditions are
satisfied: $($a$)$ $LM^*=ML^*$; $($b$)$ the restriction of $A(L,M)$ to the
subspace $J(\Lambda)$ is injective.}

\medskip

\noindent {\normalsize\bf Proof}. First of all we prove the
equivalence of the following assertions:

(1) $\Lambda\supset\Lambda^{[\bot]}$; (2) $LM^*=ML^*$.

Indeed, by definition $\Lambda=\Ker A(L,M)$; on the other hand we have
the well-known relation $\Ker A(L,M)^{\bot}=\overline{\Ran A(L,M)^*}$.
Since $\Ker A(L,M)$ is closed, condition (1) is equivalent to the condition
(3) $J(\Ker A(L,M))\supset \Ran A(L,M)^*$. Because
$A(L,M)^*\,x=(L^*x,-M^*x)$ for every $x\in\GG$, the equivalence of (2)
and (3) follows immediately.

Now let $\Lambda$ be a Lagrangian subspace, then
$J(\Lambda)=\Ker A(L,M)^{\bot}$; therefore, the
restriction of $A(L,M)$ to $J(\Lambda)$ is obviously injective. On
the other hand, if conditions (a) and (b) are satisfied, then
$J(\Lambda)\supset\Lambda^{\bot}$. Moreover, if
$J(\Lambda)\ne \Lambda^{\bot}$, then $J(\Lambda)$ contains a non-zero
element from $\Ker A(L,M)$, and we have a contradiction with (b). \halm

Note that a finite-dimensional version of Proposition B has been given in
\cite{KS}, based on different arguments.

\medskip

The self-adjoint extensions of $S$ defined by the conditions
$\Gamma^{(1)}x=0$ and
$\Gamma^{(2)}x=0$ will be denoted by $H^{(1)}$ and $H^{(2)}$, respectively;
they correspond to the Lagrangian subspaces $\{0\}\oplus\GG$ and
$\GG\oplus\{0\}$, respectively. If $\Lambda$ is the graph of a
self-adjoint operator $L$ in $\GG$ (i.e. if $\Lambda$ is transversal to
$\{0\}\oplus\GG$: $\Lambda\cap(\{0\}\oplus\GG)=\{0\}$), then the
condition (\ref{n1.3}) takes the simpler form
\begin{equation}
                                      \label{n1.5}
\Gamma^{(2)}x=L\Gamma^{(1)}x\,.
\end{equation}
The self-adjoint extension $H^\Lambda$ of $S$ is defined by a boundary
condition of the form (\ref{n1.5}) with a self-adjoint $L$, if and only
if $H^\Lambda$ is disjoint from
$H^{(1)}$ (this means that $\DD(H^{(1)})\cap \DD(H^\Lambda)=\DD(S)$).

On the other hand, at least in the case of a finite-dimensional $\GG$ we
can always define a given extension $H^\Lambda$ by a condition of the form
(\ref{n1.5}). This may be done with the help of the above mentioned Witt
theorem, but a more  useful way is to use the complex version of the\
Arnold Lemma \cite{Arn}. To state this lemma we need some auxiliary
notations. Let $\ee_1, \ldots, \ee_n$ be a fixed orthonormal basis in $\GG$,
then the vectors $\ba_j=(\ee_j,\,0)$ and $\bb_j=(0,\,\ee_j)$ (\jj) form
a symplectic basis in $\GG\oplus\GG$:
\begin{equation}
                                  \label{n1.5a}
[\ba_j|\ba_k]=[\bb_j|\bb_k]=0\, \quad [\ba_j|\bb_k]=
-[\bb_k|\ba_j]=\delta_{jk}\,.
\end{equation}
Let $\eta$ be a subset of $\{1,\ldots,n\}$,
$\eta'=\{1,\ldots,n\}\setminus\eta$; by virtue of (\ref{n1.5a})
the linear hull of the set $\{\ba_j:\,j\in\eta\}\cup\{\bb_j:\,j\in\eta'\}$,
is a Lagrangian subspace of $\GG\oplus\GG$ which is called a
{\it coordinate subspace} and denoted by $\GG_\eta$. It is clear that if
$\eta=\{1,\ldots,n\}$, then $\GG\oplus\{0\}=\GG_\eta$,
$\{0\}\oplus\GG=\GG_{\eta'}$.

\medskip

{\bf Proposition C}  (Arnold's Lemma). {\it Let $\GG$ be finite-dimensional.
Then every Lagrangian subspace of $\GG\oplus\GG$ is transversal to some
coordinate subspace}. \halm

\medskip

Moreover, $\GG\oplus\GG=\GG_\eta\oplus\GG_{\eta'}$ where the sums
are orthogonal with respect to the standard scalar product $\bra x|y\ket$
in $\GG\oplus\GG$. Denote the orthoprojection of $\GG\oplus\GG$ onto
$\GG_\eta$ by $P_\eta$; by $J^{(1)}_\eta$ we shall denote the isomorphism
of $\GG_\eta$ onto $\GG$ which takes the elements from $\GG_\eta$
of the form $\ba_j$ or $\bb_j$ to $\ee_j$, by $J_\eta^{(2)}$ we denote the
isomorphism of $\GG_{\eta'}$ onto $\GG$ which takes the elements
from $\GG_\eta$ of the form $\ba_j$ to $-\ee_j$ and of the form $\bb_j$
into $\ee_j$. Let now $(\GG,\,\Gamma)$ be a boundary value space for
a symmetric operator $S$, denote
$\Gamma_\eta^{(1)}=J^{(1)}_\eta P_\eta\Gamma$,
$\Gamma_{\eta}^{(2)}=J^{(2)}_\eta P_\eta\Gamma$. Then the triple
$\left(\GG,\,\Gamma_\eta^{(1)},\,\Gamma_\eta ^{(2)}\right)$ is a boundary
value space for $S$ as well.
For example, if $\eta=\{1,\ldots,n\}$, then $\Gamma_\eta^{(j)}=\Gamma^{(j)}$;
on the other hand, $\Gamma^{(1)}_{\emptyset}=\Gamma^{(2)}$,
$\Gamma^{(2)}_{\emptyset}=-\Gamma^{(1)}$.

By virtue of the Arnold lemma, for every Lagrangian subspace $\Lambda
\subset\GG\oplus\GG$ there exists $\eta\subset\{1,\ldots,n\}$ such that the
self-adjoint extension $H^\Lambda$ is given by the boundary condition of the
form $\Gamma^{(2)}_\eta x=L \Gamma^{(1)}_\eta x$ where $L$ is a self-adjoint
operator in $\GG$. We shall denote this extension by $H^{L,\,\eta}$; the
representation of $H^\Lambda$ in the form $H^{L,\,\eta}$ is, clearly, not
unique. The extensions of $S$ defined by the conditions
$\Gamma_\eta^{(j)}x=0$ ($j=1,2$) will be defined by $H^{(j)}_\eta$.

There exists a very convenient expression for the resolvent
$R^{\Lambda}(z)=(H^{\Lambda}-z)^{-1}$ of the operator $H^{\Lambda}$ which
is given by the so-called Krein resolvent formula. To give this formula,
we need some preliminary notions (details may be found in \cite{DM},
\cite{KL}). Let $z\in\CC\setminus\RR$, denote by $\N_z$
the deficiency subspace for $S$: $\N_z={\rm Ker}\,(S^*-z)$. It may be
proven that the restrictions of both the operators $\Gamma^{(j)}$ ($j=1,2$)
to $\N_z$ are linear-topological isomorphisms of $\N_z$ onto $\GG$; we
denote these restrictions as $\Gamma^{(j)}(z)$. Moreover, the operators
$\gamma(z)=\left(\Gamma^{(1)}(z)\right)^{-1}$ form a holomorphic family of
elements from the Banach space ${\cal L}(\GG,\,\HH)$ of all linear continuous
operators from $\GG$ to $\HH$. Further, the operators
$Q(z)=\Gamma^{(2)}\gamma(z)$ form a holomorphic family 
in the Banach space ${\cal L}(\GG,\,\GG)$. The holomorphic operator-valued
functions $z\mapsto \gamma(z)$ and $z\mapsto Q(z)$ have analytic
continuations on the set $\rho(H^{(1)})$ of the regular values of
$H^{(1)}$: $\rho(H^{(1)})=\CC\setminus\sigma(H^{(1)})$. This assertion
follows from the relations below, which are valid for every
$z,\zeta\in\CC\setminus\RR$:

\bigskip

$$
\gamma(z)=\gamma(\zeta)+(z-\zeta)(H^{(1)}-z)^{-1}\gamma(\zeta);
$$
%
%
$$
Q(z)-Q(\zeta)=(z-\zeta)\gamma^*(\bar\zeta)\gamma(z).
$$
%
%
The functions $\gamma:\rho(H^{(1)})\to {\cal L}(\GG,\,\HH)$ and
$Q:\rho(H^{(1)})\to {\cal L}(\GG,\,\GG)$ are called Krein $\Gamma$-field
and Krein $\Q$-function of the operator $S$ associated with the boundary
value space $(\GG,\Gamma)$.

Further we shall consider a subspace $\Lambda\subset\GG\oplus\GG$ as
the graph of a multi-valued linear operator
$M_{\Lambda}$ with the domain $\DD(M_{\Lambda})=P_1(\Lambda)$.
The operator $M_{\Lambda}$ takes each $x\in\DD(M_\Lambda)$ to an affine
subspace $\{y\in\GG:\,(x,y)\in\Lambda\}$ of $\GG$. For every subspace
$\Lambda\subset\GG\oplus\GG$ we denote by $\Lambda^{-1}$ the "inverse"
subspace $\Lambda^{-1}=\{(x,y)\in\GG\oplus\GG:\,(y,x)\in\Lambda\}$. In
particular, if $\Lambda$ is the graph of an invertible operator
$L:\,\DD(L)\to\GG$, then $\Lambda^{-1}$ is the graph of the inverse operator
$L^{-1}$. In the following we shall identify mappings and their graphs if
this does not lead to ambiguities.

\bigskip

{\bf Theorem D}. {\it Let $S$ be a symmetric operator in a Hilbert space
$\HH$ with boundary value space $(\GG,\,\Gamma)$, and let $\gamma$ and $Q$
be the corresponding $\Gamma$-field and $\Q$-function for $S$, respectively.
Suppose that $H^{\Lambda}$ is a self-adjoint extension of $S$ associated
with a Lagrangian subspace $\Lambda$ of $\GG\oplus\GG$. Then for every
$z\in\rho(H^{(1)})\cap\rho(H^{\Lambda})$ the subspace
$\left[Q(z)-\Lambda\right]^{-1}$ is the graph of a bounded
$($single-valued$)$ operator in $\GG$ and the resolvent of
$R^{\Lambda}(z)=(H^{\Lambda}-z)^{-1}$ has the form
\begin{equation}
                                   \label{n1.7}
R^{\Lambda}(z)=R^{(1)}(z)-\gamma(z)\left[Q(z)-\Lambda\right]^{-1}
\gamma^*(\bar z)\,,
\end{equation}
where $R^{(1)}(z)=(H^{(1)}-z)^{-1}$ is the resolvent of $H^{(1)}$}. \halm

\medskip

If $H^\Lambda$ and $H^{(1)}$ are not disjoint, then the Krein formula
(\ref{n1.7}) contains a multi-valued operator $\Lambda$. To avoid the use
of such operators we can proceed as follows \cite{KL}. Let $\Lambda$ be a
Lagrangian subspace of $\GG\oplus\GG$ and $U_\Lambda$ be its Cayley
transform. Denote by $P_\Lambda$ the orthogonal projection of $\GG$ onto
subspace $\GG_\Lambda=\overline{{\rm Ran}\,(U_\Lambda-I)}$, by $J_\Lambda$
the canonical embedding of $\GG_\Lambda$ into $\GG$,
and by $I_\Lambda$ the identity operator in $\GG_\Lambda$.
Then $V_\Lambda=P_\Lambda U_\Lambda J_\Lambda$ is a unitary operator in
$\GG_\Lambda$, and $1$ is not an eigenvalue of this operator. Therefore,
$L=i(I_\Lambda+V_\Lambda)(I_\Lambda-V_\Lambda)^{-1}$ is a self-adjoint
operator in $\GG_\Lambda$, and
\begin{equation}
                                  \label{n1.8}
[Q(z)-\Lambda]^{-1}=J_\Lambda[P_\Lambda Q(z) J_\Lambda - L]^{-1}P_\Lambda\,.
\end{equation}
Moreover, the following proposition holds \cite{KL}:

\medskip

{\bf Proposition E}. {\it Let $L_n$ be a self-adjoint operator in
$\GG$ of the form $L_n=J_\Lambda L P_\Lambda + n(I-P_\Lambda)$. Then for
every $z\in\rho(H^{(1)})\cap\rho(H^\Lambda)$
%
%
%
$$
\lim\limits_n\, [Q(z)-L_n]^{-1}=J_\Lambda[P_\Lambda Q(z)
J_\Lambda - L]^{-1}P_\Lambda\,
$$
in the strong operator topology}. \halm

\medskip

If $\GG$ is finite-dimensional, then we can
adapt  the Arnold Lemma to avoid the use of multi-valued mappings in the
Krein formula. Namely, denote the Krein $\Gamma$-field and $\Q$-function
for the boundary value space $(\GG,\Gamma^{(1)}_\eta,\Gamma^{(2)}_\eta)$ by
$\gamma_\eta(z)$ and $Q_\eta(z)$, respectively. Since $H^\Lambda$ coincides
with some operator of the form $H^{L,\,\eta}$, then
(\ref{n1.7}) may be rewritten in the form
\begin{equation}
                                   \label{n1.9a}
R^{\Lambda}(z)\equiv R^{L,\,\eta}(z)=
R^{(1)}_\eta(z)-\gamma_\eta(z)\left[Q_\eta(z)-L\right]^{-1}
\gamma_\eta^*(\bar z)\,,
\end{equation}
where $R^{(1)}_\eta(z)=(H^{(1)}_\eta-z)^{-1}$.

\vskip 5ex

\noindent{\bf 2.~Schr\"odinger operator on a "hedgehog shaped" space}

\bigskip

\noindent Consider a complete (not necessarily connected)
Riemannian manifold $X$ of dimension $d$, with metric $g_{\mu\nu}$.
We shall denote by $g$ the determinant ${\rm det}\, (g_{\mu\nu})$,
by $d\lambda$ the Riemannian measure, and
by $r(x,y)$ the geodesic distance on $X$. Fix a non-empty finite subset
$\{q_1,\,\ldots,\,q_n \}$ of $X$, and let $\Rs^{(j)}$ (\jj) be copies of
the half-line $\Rs=\{x\in\RR:\,x\ge0\}$.
Let $\hat X$ be the topological space obtained from the disjoint union
$X\sqcup\Rs^{(1)}\sqcup\ldots\sqcup\Rs^{(n)}$ by gluing
the point $0\in \Rs^{(j)}$
to the point $q_j$. The "hedgehog shaped" topological space  $\hat X$ may be
considered as a limit of manifolds with $n$ horns
as the widths of the horns tend to zero.
Let $\HH_0:=L^2(X,\,d\lambda)$, $\HH_j:=L^2(\Rs^{(j)},\,dx)$.
The sum of the Riemannian measure $d\lambda$ on $X$ and
the Lebesgue measures $dx$ on $\Rs^{(j)}$ is a natural measure $d\mu$
on $\hat X$; the space $L^2(\hat X,\,d\mu)$ will be identified with the
space $\HH:=\HH_0\oplus \HH_1\oplus\ldots\oplus\HH_n$.

To define a Schr\"odinger operator on $\hat X$ we proceed as follows.
Consider the symmetric operator $\tau$ in $\HH_0$, with domain
$C^{\infty}_0(X)$, defined by the differential expression
%
%
%
$$
\tau=-g^{-1/2}(x)\left(\dd_{\mu}+i{\cal A}{_\mu}(x)\right)
g^{1/2}(x)g^{\mu\nu}(x)\left(\dd_{\nu}+i{\cal A}{_\nu}(x)\right)+p(x)\,,
$$
%
%
where ${\cal A}_{\mu}$ ($\mu=1,\ldots,d$) and $p$ are real-valued smooth
functions on $X$ (${\cal A}_{\mu}$ are the components of a vector potential
${\cal A}$ of a certain magnetic field on $X$, and $p$ is the scalar
potential of an electric field). We denote the closure of $\tau$ in
$\HH_0$ by $H_0$, and suppose that the potentials ${\cal A}$ and $p$ are
chosen in such a way that $H_0$ is a self-adjoint operator in $\HH_0$,
i.e. we assume that $\tau$ is essentially self-adjoint.
Note that this is the case, if ${\cal A}_{\mu}$ and $p$ have compact
supports, in particular, if $X$ is compact. If ${\cal A}=0$ and $p=0$ we
get the Laplace--Beltrami operator $-\Delta_X$  on $X$. To use the
techniques of the operator extension theory we need the condition

\medskip

\noindent (C) {\it $\DD(H_0)$ imbeds in $C(X)$}.

\medskip

\noindent By virtue of the well-known Sobolev embedding theorems, this
condition is satisfied if and only if $d\le 3$.
Therefore, {\bf from this point on we suppose that} $0<d\le 3$.

Let now $S_0$ be the operator in $\HH_0$ that is the restriction of $H_0$
to the domain
%
%
%
$$
\DD(S_0)=\{f\in\DD(H_0):\, f(q_j)=0\quad \forall\, j=1,\ldots,\,n \}\,.
$$
%
%
It is clear that $S_0$ is symmetric in $\HH_0$, and it is easy to prove
that the deficiency indices of $S_0$ are $(n,\,n)$. Denote next by
$S_j$ (\jj) the closure in $\HH_j$ of the operator $-d^2/dx^2$ defined on
$C_0^{\infty}(0,\infty)$; $S_j$ is a symmetric operator in $\HH_j$ with the
deficiency indices $(1,1)$. Finally, we set
$S:=S_0\oplus S_1\oplus\,\ldots\,\oplus S_n$; it is evident that
$S$ is a symmetric operator in $\HH$ with deficiency indices $(2n,\,2n)$.

\medskip

\noindent{\bf Definition}. Any self-adjoint extension $H$ of the operator
$S$ we shall call a {\it Schr\"odinger operator on $\hat X$}
with vector potential ${\cal A}$ and scalar potential $p$.

\medskip

According to the theory presented in Sec.~1,
to describe all the Schr\"odinger operators on $\hat X$ with given vector and
scalar potentials we must construct a boundary value space for $S$. For this
purpose we construct boundary value spaces for the operators $S_0$, $S_1$,
.... , $S_n$ and take the direct sum of these spaces. Let us start with a
simple case of the operators $S_j$ (\jj).

\medskip

{\bf Lemma 1}. {\it Set $\GG_j=\CC$ and define the operators
$\Gamma_j^{(1)},\,\Gamma_j^{(2)}\in {\cal L}(\DD(S_j^*),\GG_j)$, \jj,
by the rule:
\begin{equation}
                              \label{n2.3}
\Gamma_j^{(1)}(f)=-f'(0)\,,\quad \Gamma_j^{(2)}(f)=f(0)\,.
\end{equation}
Then the triple $(\GG_j,\Gamma_j^{(1)},\Gamma_j^{(2)})$ is a  boundary
value space for $S_j$}.

\medskip

\noindent We omit the simple proof. 

It is clear that $H^{{\rm N}}_j:=H_j^{(1)}$ and $H^{{\rm D}}_j:=H_j^{(2)}$
is the free Schr\"odinger operator on the semi-axis $\Rs^{(j)}$
with the Neumann and Dirichlet boundary condition at the point $x=0$,
respectively. Since the space $\GG_j$ is one-dimensional, the corresponding
$\Gamma$-field $\gamma_j(z)$ may be considered as a holomorphic function on
$\CC\setminus\Rs=\rho(H_j^{(1)})$ with values in
$\HH_j=L^2(\Rs)$, and the corresponding $\Q$-function $Q_j(z)$ as a
holomorphic function in $\CC\setminus\Rs$. It is clear that $\gamma_j$ and
$Q_j$ are independent of $j$.

\medskip

{\bf Lemma 2}. {\it The $\Gamma$-field and the $\Q$-function for $S_j$
associated with the boundary value space
$\left(\GG_j\,,\Gamma_j^{(1)},\,\Gamma_j^{(2)}\right)$ are given by}:
$$
\gamma_j(z)(x)=\frac{1}{\sqrt{-z}}\exp(-\sqrt{-z}\,x)\,,
$$
\begin{equation}
                                 \label{n2.4}
Q_j(z)=\frac{1}{\sqrt{-z}}\,.
\end{equation}

\medskip

\noindent{\bf Remark 2}. Throughout the paper, the continuous branch of
the square root is chosen in $\CC\setminus(-\infty,0)$, such that
${\rm Re}\,\sqrt{z}>0$ if $z\ne0$.

\medskip

\noindent {\normalsize\bf Proof}.
It is clear that $S_j$ is the restriction of $H_j^{{\rm N}}$ to
the domain $\{f\in\DD(H_j^{{\rm N}})\,:f(0)=0\}$. On the other hand, it
is easy to check that the Green function $G_j^{\rm N}(x,y;z)$ of
the Neumann operator $H_j^{{\rm N}}$ is given as
\begin{equation}
                               \label{n2.5}
G_j^{\rm N}(x,y;z)=
\frac{1}{2\sqrt{-z}}\left[\exp(-\sqrt{-z}\,|x-y|)+\exp(-\sqrt{-z}\,(x+y))
\right]\,.
\end{equation}
Hence, the function $g(x)=G^{\rm N}_j(x,0;z)$ is a non-zero element of
$\Kr(S_j^*-z)$. Since $-g'(0)=1$, the operator $\Gamma_j^{(1)}\gamma_j(z)$
is the identity on $\GG_j$. Therefore, $\gamma_j(z)$ is the $\Gamma$-field.
The equation $Q_j(z)=\Gamma_j^{(2)}\gamma(z)$ is trivial, so $Q_j(z)$ is
the $\Q$-function. \halm

Sometimes it is more convenient to use the boundary value space
$(\GG_j,\Gamma^{(1)}_{\emptyset,\,j},\Gamma^{(2)}_{\emptyset,\,j})$ (see
notations in the preceding section). It is clear that
$\Gamma_{\emptyset,\,j}^{(1)}f=f(0)$,  and
$\Gamma_{\emptyset,\,j}^{(2)}f=f'(0)$; thus
$H_{\emptyset,\,j}^{(1)}=H_j^{{\rm D}}$,
$H_{\emptyset,\,j}^{(2)}=H_j^{{\rm N}}$.  Using the definitions
and Lemma 2, we get

\medskip

{\bf Lemma 3}. {\it The $\Gamma$-field $\gamma_j^{\rm D}$ and $\Q$-function
$Q_j^{\rm D}$ for $S_j$ associated with the boundary value space
$\left(\GG_j\,,\Gamma_{\emptyset,\,j}^{(1)},\,\Gamma_{\emptyset,\,j}^{(2)}
\right)$ have the following form}:
$$
\gamma_j^{\rm D}(z)(x)\equiv \gamma_{\emptyset,\,j}(z)(x)=
\exp(-\sqrt{-z}\,x)\,,
$$
\begin{equation}
                                 \label{n2.6}
Q_j^{\rm D}(z)\equiv Q_{\emptyset,\,j}(z)=-\sqrt{-z}\,.
\end{equation}

\medskip

Now we turn to the operator $S_0$. First of all, denote by $R_0(z)$ the
resolvent for $H_0$, $R_0(z)=(H_0-z)^{-1}$; by $G_0(x,y;z)$ we shall denote
the Green function for $H_0$ (the integral kernel of $R_0(z)$ in the space
$L^2(X,\,d\lambda)$). Fix $q\in X$ and $z\in\rho(H_0)$, then near $q$ the
function $G_0(x,q;z)$ has the expansion  \cite{Avr}, \cite{CdV}, \cite{Mir}:
\begin{equation}
                                  \label{n2.7}
G_0(x,q;z)=F_0(x,q)+F_1(x,q;z)+R(x,q;z)\,,
\end{equation}
where $F_0$ is independent of the spectral parameter $z$ and has the
following form:
\begin{equation}
                                \label{n2.8}
F_0(x,q)=\cases{\displaystyle
-\frac{c_1(x,q)}{2}r(x,q)\,, &if $d=1$;\cr
\noalign{\medskip}
\displaystyle-\frac{c_2(x,q)}{2\pi} \ln r(x,q)\,, & if  $d=2$;\cr
\noalign{\medskip}
\displaystyle\frac{c_3(x,q)}{4\pi} [r(x,y)]^{-1}\,, & if $d=3$.\cr}
\end{equation}
Here $c_j(x,q)$ ($j=1,2,3$) does not depend on $z$, is a continuous
functions of $x$, and $c_j(q,q)=1$; moreover, $c_1$ is a smooth function
of $x$.
Further, the function $F_1$ is continuous with respect to $x$; as for the
remainder term $R$, it has the following behavior near $q$ as $x\to q$:
\begin{equation}
                                \label{n2.9}
R(x,q;z)=\cases{\displaystyle
o(r(x,q))\,, &if $d=1$;\cr
\noalign{\medskip}
\displaystyle o\,(1)\,, & if  $d=2$ or $d=3$\,.\cr}
\end{equation}
Finally, $F_1$ and $R$ are analytic functions of $z$ in the domain
$\rho(H_0)=\CC\setminus \sigma(H_0)$.

For $z\in \rho(H_0)$ define a matrix $Q_0(z)$ by the relations:
\begin{equation}
                               \label{n2.10}
[Q_0(z)]_{lm}:=\cases{
G_0(q_l,q_m;z)\,, & if $l\ne m$;\cr
\noalign{\medskip}
F_1(q_l,q_l;z)\,, & if $l=m$\cr}
\end{equation}
(note that $[Q_0(z)]_{lm}=G_0(q_l,q_m;z)$ for all $l$ and $m$, if $d=1$).
Clearly,  $Q_0(z)$ is a holomorphic matrix-valued function in the domain
$\rho(H_0)$ obeying the condition
\begin{equation}
                           \label{n2.11}
\overline{[Q_0(z)]}_{lm}=[Q_0(\bar z)]_{ml}\,.
\end{equation}
The following assertion is needed below (see  \cite{Koc2}, \cite{Zor},
\cite{GMC}).

\medskip

{\bf Lemma 4.} {\it If $z\in\CC\setminus\RR$, then the
functions $G_0(\cdot\,,q_j;z)$, \jj{}, form a vector basis in the deficiency
subspace $\N_z=\Kr(S_0^*-z)$}. 

\medskip

Fix $z\in\CC\setminus\RR$, then
$\DD(S_0^*)=\DD(\bar S_0)\dot+\N_z\dot+\N_{\bar z}$ (there is an algebraic
direct sum here; if $z=i$, then this sum is orthogonal with respect to the
scalar product $\bra x|y\ket_S$). By Lemma 4 each function $f$ from
$\DD(S_0^*)$ has the following asymptotic expansion near the point $q_j$:

\begin{equation}
                            \label{n2.12}
f(x)=a_j(f)F_0(x,q_j)+b_j(f)+R(x)\,,
\end{equation}
where $a_j(f),\,b_j(f)\in \CC$, and the behavior of the remainder term
$R(x)$ is given by (\ref{n2.9}) as $x\to q_j$.

\medskip

{\bf Lemma 5}. {\it  Set $\GG_0=\CC^n$ and  define operators
$\Gamma_0^{(1)},\,\Gamma_0^{(2)}\in {\cal L}\left(\DD(S_0^*),\GG_0\right)$ by
$$
\Gamma_0^{(1)}(f):=\left(a_j(f)\right)_{1\le j \le n}\,,
$$
%
%
$$
\Gamma_0^{(2)}(f):=\left(b_j(f)\right)_{1\le j \le n}\,,
$$
%
where $a_j(f)$ and $b_j(f)$ are the constants from $(\ref{n2.12})$.
Then the triple $\left(\GG_0,\Gamma_0^{(1)},\Gamma_0^{(2)}\right)$ is a
space of boundary values for $S_0$.}

\medskip

\noindent {\normalsize\bf Proof}. Since $C^{\infty}_0(X)\subset \DD(H_0)$ it
is easy to conclude that the mapping $f\mapsto\left(\Gamma^{(1)}_0f,\,
\Gamma^{(2)}_0f\right)$ is surjective. It remains to prove
the condition (\ref{n1.2}).

Consider the sesquilinear forms $B_1$, $B_2$ defined as follows:
$$
B_1(f,g):=\bra f|S_0^*g\ket-\bra S_0^*f|g\ket,
$$
$$
B_2(f,g):=\bra \Gamma_0^{(1)}f|\Gamma_0^{(2)}g\ket-
\bra \Gamma_0^{(2)}f|\Gamma_0^{(1)}g\ket=\sum\limits_{j=1}^n
\left[\overline{a_j(f)}b_j(g)-\overline{b_j(f)}a_j(g)\right],
$$
and set
%
%
$$
g^{\pm}_j(x)=G_0(x,q_j;\pm i),\quad j=1,\ldots,n.
$$
%
It is easy to check the following properties of the functions
$g^{\pm}_j$:
\begin{equation}
                         \label{n2.14}
({\rm i})\quad S_0^*g^{\pm}_j=\pm ig^{\pm}_j; \qquad
({\rm ii})\quad a_j(g^{\pm}_k)=\delta_{jk}; \qquad
({\rm iii})\quad b_j(g^{\pm}_k)=Q_0^{jk}(\pm i).
\end{equation}
To prove the lemma, it is enough to verify that $B_1(f,g)=B_2(f,g)$ for
$f,g\in \DD(S_0^*)$. Since $\DD(S_0^*)=\DD(\bar S_0)\oplus\N_i\oplus\N_{-i}$,
it is enough to check the equality $B_1(f,g)=B_2(f,g)$ for all functions
$f,g\in \DD(S_0)\cup\{g^{\pm}_j:\,j=1,\ldots,n\}$.
It is clear that $a_j(f)=b_j(f)=0$ if $f\in \DD(S_0)$; therefore
$B_1(f,g)=0=B_2(f,g)$ if $f\in \DD(S_0)$ or $g\in \DD(S_0)$.
By (i) from (\ref{n2.14}), $B_1(g^+_j,g^-_k)=0$
$\forall\,j,k\in\{1,\ldots,n\}$. On the other hand, Eqs.~(ii) and (iii) from
(\ref{n2.14}) and (\ref{n2.11}) imply that
\begin{equation}
                       \label{n2.15}
B_2(g^+_j,g^-_k)=b_j(g^-_k)-\overline{b_k(g^+_j)}=
[Q_0(-i)]_{jk}-\overline{[Q_0(i)]}_{kj}=0.
\end{equation}
Hence, $B_1(g^+_j,g^-_k)=B_2(g^+_j,g^-_k)$ $\forall\,j,k\in\{1,\ldots,n\}$.
Since $B_l(f,g)=-\overline{B_l(g,f)}$ ($l=1,2$),
we have:
$B_1(g^-_j,g^+_k)=B_2(g^-_j,g^+_k)$ $\forall\,j,k\in\{1,\ldots,n\}$.
Similarly, we get
%
%
$$
B_2(g^+_j,g^+_k)=b_j(g^+_k)-\overline{b_k(g^+_j)}=
[Q(i)]_{jk}-\overline{[Q(i)]}_{kj}=[Q(i)]_{jk}-[Q(-i)]_{jk}.
$$
%
%
Further
$$
B_1(g^+_j,g^+_k)=2i\bra g^+_j|g^+_k\ket=
$$
\begin{equation}
                            \label{n2.17}
\displaystyle
2i\int\limits_X\overline{G_0(x,q_j;i)}G_0(x,q_k;i)\,d\lambda(x)=
2i\int\limits_X G_0(q_j,x;-i)G_0(x,q_k;i)\,d\lambda(x)\,.
\end{equation}
Using the Hilbert resolvent identity we obtain from (\ref{n2.17})
in case $j\ne k$:
%
%
$$
B_1(g^+_j,g^+_k)=G_0(q_j,q_k;i)-G_0(q_j,q_k;-i)=B_2(g^+_j,g^+_k).
$$
%
%
If $j=k$, then using the Hilbert identity again we get
$$
\displaystyle
B_1(g^+_j,g^+_j)=
2i\lim\limits_{q\to q_j}\int\limits_X
G_0(q,x;-i)G_0(x,q_j;i)\,d\lambda(x)=
$$
$$
\lim\limits_{q\to q_j}\left[
G_0(q,q_j;i)-G_0(q,q_j;-i)\right]=
$$
%
$$
[Q(i)]_{jj}-[Q(-i)]_{jj}=B_2(g^+_j,g^+_j)
$$
%
(of course, in the case $d=1$ we can omit the limiting procedure).
The proof of the equalities $B_1(g^-_j,g^-_k)=B_2(g^-_j,g^-_k)$
is similar. \halm

\medskip

\noindent{\bf Remark 3}. It is clear that in the case $d=1$ we have
$b_j(f)=f(q_j)$. Moreover, we can get a simple expression for $a_j(f)$ in
this case. Namely, choose a chart $U\subset X$ such that $q_j\in U$ $\forall$
\jj\quad and $U$ is isometric to an interval $(a,b)\subset\RR$. Using the
Cartesian coordinates in $U$ we obtain from (\ref{n2.12}) that every function
$f\in\DD(S_0^*)$ has the following expansion near each point $q_j$\,:
\begin{equation}
                          \label{n2.20}
f(x)=-\frac{1}{2}a_j(f)|x-q_j|+f(q_j)+o(|x-q_j|),
\end{equation}
thus it follows from (\ref{n2.20}) that
%
%
%
$$
a_j(f)=f'(q_j-0)-f'(q_j+0)\,,
$$
%
where the derivative is taken with respect to  the Cartesian coordinate in
$U$.

\medskip

Now we describe the Krein $\Gamma$-field and $\Q$-function for $S_0$
associated with the boundary value space
$\left(\GG_0,\Gamma_0^{(1)}, \Gamma_0^{(2)}\right)$.

\medskip

{\bf Lemma 6}. {\it The  Krein $\Gamma$-field for $S_0$ associated with
the boundary value space $\left(\GG_0,\Gamma_0^{(1)}, \Gamma_0^{(2)}\right)$
is an operator valued family $\gamma_0(z)\in {\cal L}(\GG_0,\HH_0)$ defined
for an element $\zeta=(\zeta_j)_{1\le j \le n}$ from $\GG_0=\CC^n$ by
\begin{equation}
                                   \label{n2.22}
\gamma_0(z)(\zeta)=\sum\limits_{j=1}^n \zeta_j G_0(\cdot,q_j;z)\,.
\end{equation}
The corresponding $\Q$-function coincides with the matrix-valued function
$Q_0(z)$.}

\medskip

\noindent {\normalsize\bf Proof}. To prove the first part of the lemma, it is
enough to check that  $\Gamma_0^{(1)}\gamma_0(z)$ is the identity operator on
$\GG_0$, but this follows immediately from the definition of
$\Gamma_0^{(1)}$ and from (\ref{n2.7}) and (\ref{n2.8}).

\medskip

Let $g_k(x)=G_0(x,q_k;z)$, then $b_j(g_k)=[Q_0(z)]_{jk}$ by definition.
Thus for $\zeta\in\GG_0$ we have
$$
\left[\Gamma_0^{(2)}\gamma_0(z)\zeta\right]_j=
\sum\limits_{k=1}^n [Q_0(z)]_{jk}\zeta_k\,;
$$
therefore, $Q_0(z)$ is the $\Q$-function. \halm

\medskip

Now we set
\begin{eqnarray}
                           \label{n2.23}
\GG&:=&\GG_0\oplus\GG_1\oplus\ldots\oplus\GG_n\,\,(=\CC^{2n});
\nonumber\\
\Gamma^{(j)}&:=&
\Gamma_0^{(j)}\oplus\Gamma_1^{(j)}\oplus\ldots\oplus\Gamma_n^{(j)}
\quad(j=1,2)\,;\nonumber\\
\gamma(z)&:=&\gamma_0(z)\oplus\gamma_1(z)\oplus\ldots\oplus\gamma_n(z)
\quad(z\in\CC\setminus\RR)\,;\nonumber\\
Q(z)&:=&
Q_0(z)\oplus Q_1(z)\oplus\ldots\oplus Q_n(z)\quad(z\in\CC\setminus\RR)\,.
\end{eqnarray}

Then the following theorem is an evident consequence of the preceding lemmas.

\medskip

{\bf Theorem 1}. {\it The triple $\left(\GG,\Gamma^{(1)},\Gamma^{(2)}\right)$
is a boundary value space for the operator $S$. The corresponding Krein
$\Gamma$-field and $\Q$-function coincide with $\gamma(z)$ and $Q(z)$,
respectively.
The operator $H=H^{(1)}$ given by the boundary condition $\Gamma^{(1)}f=0$
coincides with the  direct sum $H=H_0\oplus H^{{\rm N}}_1\oplus\ldots\oplus
H^{{\rm N}}_n$ $($we shall denote this operator by $H_{{\rm N}}$}). \halm

\medskip

\noindent{\bf Remark 4}. It is convenient to describe explicitly the
boundary value space $(\GG, \Gamma^{(1)}_\eta, \Gamma^{(2)}_\eta)$ for an
arbitrary set $\eta\subset\{1,\ldots,2n\}$. Denote
$$
\theta=\eta\cap\{1,\ldots,n\}\,,\quad  \omega=\eta\cap\{n+1,\ldots,2n\}\,,
$$
\begin{equation}
                                   \label{n2.23a}
\theta'=\{1,\ldots,n\}\setminus\eta\,,\quad  \omega'=\{n+1,\ldots,2n\}
\setminus\eta\,.
\end{equation}
Then
\begin{eqnarray}
                          \label{n2.23b}
\Gamma_{\eta}^{(l)}&=&\Gamma_{\theta,\,0}^{(l)}\oplus
\tilde\Gamma_1^{(l)}\oplus\ldots\oplus\tilde\Gamma_n^{(l)}\quad(l=1,2)\,;
\nonumber\\
\gamma_\eta(z)&=&\gamma_{\theta,\,0}(z)\oplus\tilde\gamma_1(z)\oplus
\ldots\oplus \tilde\gamma_n(z)\quad(z\in\CC\setminus\RR)\,;\nonumber\\
Q_\eta(z)&=&Q_{\theta,\,0}(z)\oplus \tilde Q_1(z)\oplus\ldots\oplus
\tilde Q_n(z)\quad(z\in\CC\setminus\RR)\,.
\end{eqnarray}
Here for $j=1,\ldots,n$
%
$$
\tilde\Gamma^{(l)}_j=
\cases{\Gamma^{(l)}_j\,, &if $j+n\in\omega$,\cr
\Gamma^{(l)}_{\emptyset,\,j}\,, &if $j+n\in\omega'$;\cr}
$$
%
and similarly for $\tilde\gamma_\eta(z)$ and $\tilde Q_\eta(z)$.
In particular, if $\eta=\{1,\ldots,n\}$, then we denote
$\Gamma^{(l)}_\eta=\Gamma^{(l)}_{\rm D}$,
$\gamma_\eta=\gamma_{\rm D}$, $Q_\eta=Q_{\rm D}$. The operator
$H= H^{(1)}_{\rm D}$ given by the boundary condition
$\Gamma^{(1)}_{\rm D} f=0$
coincides with the  direct sum
$H=H_0\oplus H^{{\rm D}}_1\oplus\ldots\oplus H^{{\rm D}}_n$
and will be denoted by $H_{{\rm D}}$.

\medskip

Now we can describe all Schr\"odinger operators on $\hat X$ with given
vector and scalar potentials in terms of boundary conditions at the points
$q_1,\ldots,q_n$. First of all, we describe the elements of
$\DD(S^*)$ as functions on $\hat X$. For $f\in L^2(\hat X)=\HH$ we denote by
$f_0,f_1,\ldots,f_n$ the components of $f$ in $L^2(X)=\HH_0$,
$L^2(\Rs^{(1)})=\HH_1$, $\ldots$ , $L^2(\Rs^{(n)})=\HH_n$, respectively.
It is clear that $f\in \DD(S^*)$ if and only if $f_j\in H^2(\Rs)$ (\jj)
whereas $f_0\in H^2_{\rm loc}(X\setminus\{q_1,\ldots,q_n\})$ and has the
asymptotics (\ref{n2.12}) near each point $q_j$.

\medskip

{\bf Theorem 2}. {\it The Schr\"odinger operators on $\hat X$ with a given
vector potential ${\cal A}$ and a given scalar potential $p$ are in bijective
correspondence with the Lagrangian subspaces of $\GG\oplus\GG$. More
precisely, if $\Lambda$ is such a subspace and $U_\Lambda$ is the Cayley
transform of $\Lambda$ having the matrix $(u_{jk})$ in the standard basis
of $\GG$, then the corresponding Schr\"odinger operator $H=H^\Lambda$ is
defined on those functions $f\in\DD(S^*)$ the components of which obey the
boundary conditions}
$$
\sum\limits_{k=1}^n\left[(\delta_{jk}-u_{jk})b_k(f_0)+
(\delta_{j,k+n}-u_{j,k+n})f_k(0)\right]=
$$
\begin{equation}
                               \label{n2.24}
i\sum\limits_{k=1}^n\left[(\delta_{jk}+u_{jk})a_k(f_0)-
(\delta_{j,k+n}+u_{j,k+n})f'_k(0)\right]\,,\quad j=1,\ldots,2n\,.
\end{equation}
{\it If $\Lambda$ is the graph of a self-adjoint operator $L$ in $\GG$ with
a Hermitian $2n\times 2n$-matrix $(\lambda_{jk})$ then conditions
$(\ref{n2.24})$ take a simpler form:}
\begin{eqnarray}
                               \label{n2.25}
b_j(f_0)&=&\sum\limits_{k=1}^n\left[\lambda_{j,k}a_k(f_0)-
\lambda_{j,k+n}f'_k(0)\right]\,,\nonumber\\
f_j(0)&=&\sum\limits_{k=1}^n\left[\lambda_{j+n,k}a_k(f_0)-
\lambda_{j+n,k+n}f'_k(0)\right]\,,
\quad j=1,\ldots,n\,.
\end{eqnarray}
{\it In the general case there are a finite subset
$\eta\subset\{1,\ldots, 2n\}$ and a Hermitian $2n\times 2n$-matrix
$L=(\lambda_{jk})$ such that the conditions
$(\ref{n2.24})$ take the following equivalent form:}
$$
\displaylines{
b_j(f_0)=\sum\limits_{k\in\theta}\lambda_{jk}a_k(f_0)-
   \sum\limits_{k\in\theta'}\lambda_{jk}b_k(f_0)-
\sum\limits_{k\in\omega}\lambda_{jk}f'_{k-n}(0)+
\sum\limits_{k\in\omega'}\lambda_{jk}f_{k-n}(0)\,,\quad j\in\theta\,;\hfill
\cr}
$$

$$
\displaylines{
a_j(f_0)=\sum\limits_{k\in\theta}\lambda_{jk}a_k(f_0)-
 \sum\limits_{k\in\theta'}\lambda_{jk}b_k(f_0)-
\sum\limits_{k\in\omega}\lambda_{jk}f'_{k-n}(0)+
\sum\limits_{k\in\omega'}\lambda_{jk}f_{k-n}(0)\,,\quad j\in\theta'\,;\hfill
\cr}
$$

$$
f_j(0)=\sum\limits_{k\in\theta}\lambda_{j+n,k}a_k(f_0)-
\sum\limits_{k\in\theta'}\lambda_{j+n,k}b_k(f_0)-
\sum\limits_{k\in\omega}\lambda_{j+n,k}f'_{k-n}(0)+
\sum\limits_{k\in\omega'}\lambda_{j+n,k}f_{k-n}(0)\,,\quad j+n\in\omega\,;
$$

\begin{equation}
                         \label{n2.26}
f'_j(0)=\sum\limits_{k\in\theta}\lambda_{j+n,k}a_k(f_0)-
\sum\limits_{k\in\theta'}\lambda_{j+n,k}b_k(f_0)-
%
%
%
\sum\limits_{k\in\omega}\lambda_{j+n,k}f'_{k-n}(0)+
\sum\limits_{k\in\omega'}\lambda_{j+n,k}f_{k-n}(0)\,,\quad j+n\in\omega'\,;
\end{equation}
{\it where the sets $\theta$, $\theta'$, $\omega$, and $\omega'$ are
defined in Remark} 4.
\medskip

\noindent {\normalsize\bf Proof}. The result follows immediately from
Theorem A, Theorem 1, and Proposition C. \halm

\medskip

Below we collect the most interesting particular cases of Schr\"odinger
operators on $\hat X$ with given potentials. For this purpose we need some
notions concerning point perturbations of Schr\"odinger operators on the
manifold $X$.
Let $B=(\beta_{jk})$ be a Hermitian $n\times n$-matrix, $\theta$ a subset of
$\{1,\ldots,n\}$, and $\theta'=\{1,\ldots,n\}\setminus \theta$. Then the
conditions
$$
f_0\in\DD(S^*_0)\,;
$$
$$
b_j(f_0)=\sum\limits_{k\in\theta}\beta_{jk}a_k(f_0)-\sum\limits_{k\in\theta'}
\beta_{jk}b_k(f_0)\,,\quad j\in\theta\,;
$$
\begin{equation}
                                    \label{n2.27}
a_j(f_0)=\sum\limits_{k\in\theta}\beta_{jk}a_k(f_0)-
\sum\limits_{k\in\theta'}\beta_{jk}b_k(f_0)\,,\quad j\in\theta'\,;
\end{equation}
define a generic self-adjoint extension $H_0^{B,\,\theta}$of the operator
$S_0$. In particular, if $B=0$ and $\theta=\emptyset$, then
$H_0^{B,\,\theta}$ is the Schr\"odinger operator $H_0$. If
$\theta=\{1,\ldots,n\}$, then the operator $H_0^B=H_0^{B,\,\theta}$ is called
a {\it point perturbation} of $H_0$ supported by the points
$q_1,\ldots,q_n$ \cite{AGHH}. Generally speaking, this perturbation
is non-local in the sense of \cite{DG}. If $B$ is a diagonal matrix,
$\beta_{jk}=\beta_j\delta_{jk}$, $\beta_j\in\RR$, then $H_0^B$ is called
a {\it local} point perturbation of $H_0$.

\medskip

In what follows we shall represent {\bf an arbitrary Hermitian
$2n\times2n$-matrix $L=(\lambda_{jk})$ in block form}:
\begin{equation}
                              \label{n2.28}
L=\left[
\begin{array}{cc}
B&A\\
A^*&C
\end{array}
\right]\,,
\end{equation}
where $B=(\beta_{jk})$ and $C=(\gamma_{jk})$ are Hermitian
$n\times n$-matrices whereas $A=(\alpha_{jk})$ is an arbitrary complex
$n\times n$-matrix.

\medskip

\noindent{\bf Examples}. We list four important particular cases of the
Schr\"odinger operator $H$.

\medskip

\noindent(1) Let $\eta=\emptyset$.
Then the conditions (\ref{n2.26}) take the following simpler form:
\begin{eqnarray}
a_j(f_0)&=&-\sum\limits_{k=1}^n\beta_{jk}b_k(f_0)+
\sum\limits_{k=1}^{n}\alpha_{jk}f_{k}(0)\,,\nonumber\\
f'_j(0)&=&-\sum\limits_{k=1}^n\bar\alpha_{kj}b_k(f_0)+
\sum\limits_{k=1}^{n}\gamma_{jk}f_k(0)\,,\quad j=1,\ldots,n\,.\nonumber
\end{eqnarray}
If $A=C=0$, then
$H=H_0^{B,\,\emptyset}\oplus H_1^{\rm N}\oplus\ldots\oplus H_n^{\rm N}$.
If, in addition, $B=0$, then
%
%
%
$$
a_j(f_0)=0\,,\quad f'_j(0)=0\,,\quad j=1,\ldots,n\,,
$$
hence $H$ coincides with $H_{{\rm N}}$.

\medskip

\noindent(2) Let $\eta=\{1,\ldots,2n\}$,
then the conditions (\ref{n2.26}) take the form
\begin{eqnarray}
b_j(f_0)&=&
\sum\limits_{k=1}^n\beta_{jk}a_k(f_0)-
\sum\limits_{k=1}^{n}\alpha_{jk}f'_{k}(0)\,,\nonumber\\
f_j(0)&=&\sum\limits_{k=1}^n\bar\alpha_{kj}a_k(f_0)-
\sum\limits_{k=1}^{n}\gamma_{jk}f'_k(0)\,,\quad j=1,\ldots,n\,,\nonumber
\end{eqnarray}
and we return to the conditions (\ref{n2.25}). We shall denote this
operator by $H^L_{\rm D}$ and call it a {\it Schr\"odinger operator of
Dirichlet type}. It is clear that this operator is disjoint from $H_{\rm N}$.
If $A=C=0$, then
$H^L_{\rm D}=S_0^{B}\oplus H_1^{\rm D}\oplus\ldots\oplus H_n^{\rm D}$.

\medskip

\noindent(3) Let $\eta=\{n+1,\ldots,2n\}$.
Then the conditions (\ref{n2.26}) become
\begin{eqnarray}
a_j(f_0)&=&-\sum\limits_{k=1}^n\beta_{jk}b_k(f_0)-
\sum\limits_{k=1}^{n}\alpha_{jk}f'_{k}(0)\,,\nonumber\\
f_j(0)&=&-\sum\limits_{k=1}^n\bar\alpha_{kj}b_k(f_0)-
\sum\limits_{k=1}^{n}\gamma_{jk}f'_k(0)\,,\quad j=1,\ldots,n\,.\nonumber
\end{eqnarray}
If $A=C=0$, then we get an operator
$H=H_0^{B,\,\emptyset}\oplus H_1^{\rm D}\oplus\ldots\oplus H_n^{\rm D}$.
If, in addition, $B=0$, then
%
%
%
%
$$
a_j(f_0)=0\,,\quad f_j(0)=0\,,\quad j=1,\ldots,n\,,
$$
%
i.e.   $H$ coincides with $H_{{\rm D}}$.

\medskip

\noindent(4) Let $\eta=\{1,\ldots,n\}$.
Then the conditions (\ref{n2.26}) take the form
\begin{eqnarray}
b_j(f_0)&=&\sum\limits_{k=1}^n\beta_{jk}a_k(f_0)+
\sum\limits_{k=1}^{n}\alpha_{jk}f_{k}(0)\,,\nonumber\\
f'_j(0)&=&\sum\limits_{k=1}^n\bar\alpha_{kj}a_k(f_0)+
\sum\limits_{k=1}^{n}\gamma_{jk}f_k(0)\,,\quad j=1,\ldots,n\,.\nonumber
\end{eqnarray}
We shall denote this operator by
$H_{\rm N}^L$  and call it a {\it Schr\"odinger operator of Neumann type}.
It is clear that this operator is disjoint from $H_{\rm D}$.
If $A=C=0$, then
$H^L_{\rm N}=H_0^{B}\oplus H_1^{\rm N}\oplus\ldots\oplus H_n^{\rm N}$. In
the case $n=2$, the operator $H^L_{\rm N}$ has been
considered in \cite{ETV} and \cite{Kis}.

\medskip

Theorem 1 implies the following description of the resolvents of
Schr\"o\-dinger operators.

\medskip

{\bf Theorem 3}. {\it Let $\Lambda$ be a Lagrangian subspace of
$\GG\oplus\GG$ and $H=H^\Lambda$
the Schr\"odinger operator defined by the boundary condition
$\Gamma f\in\Lambda$. Then the resolvent $R(z)=(H-z)^{-1}$ of $H$ is given
by the Krein formula
\begin{equation}
                               \label{n2.35}
R(z)=R_{\rm N}(z)-\gamma(z)\left[Q(z)-\Lambda\right]^{-1}\gamma^*(\bar z)\,,
\end{equation}
where $R_{\rm N}(z)=(H_{\rm N}-z)^{-1}$.

Similarly, if $H$ is defined by the boundary condition
$\Gamma_{\rm D} f\in\Lambda$, then the
resolvent $R(z)=(H-z)^{-1}$ is given by the expression
\begin{equation}
                               \label{n2.36}
R(z)=R_{\rm D}(z)-\gamma_{\rm D}(z)\left[Q_{\rm D}(z)-\Lambda\right]^{-1}
\gamma_{\rm D}^*(\bar z)\,,
\end{equation}
where $R_{\rm D}(z)=(H_{\rm D}-z)^{-1}$.
In particular, if $H=H^L_{\rm D}$  $($respectively, $H=H^L_{\rm N})$,
then} (\ref{n2.35})
({\it respectively} (\ref{n2.36})) {\it contains a single-valued operator
$\Lambda$ with  matrix $L$. In any case, using }(\ref{n1.9a}),
{\it  we can rewrite} (\ref{n2.35}) ({\it or}
(\ref{n2.36})) {\it in the form
%
%
$$
R^{L,\eta}(z)=R_{\eta}(z)-
\gamma_\eta(z)\left[ Q_\eta(z)-L\right]^{-1}\gamma_\eta^*(\bar z)\,,
$$
%
where $L$ is a Hermitian operator in $\GG$}. \halm

\vskip 5ex

\noindent{\bf 3.~Spectral  and scattering properties of the Schr\"odinger
operator on a "hedgehog shaped" space}

\bigskip

\noindent From this section on we {\bf suppose that the manifold $X$ is
compact}. Therefore, the spectrum $\sigma(H_0)$ is discrete; let
$\mu_0<\mu_1<\ldots<\mu_m<\ldots$ be the complete set of eigenvalues of
$H_0$. We shall denote the eigenspace of $H_0$ corresponding to $\mu_m$ by
$\HH_0(\mu_m)$; in  each $\HH_0(\mu_m)$ we fix an orthonormal basis
$\psi_m^{(1)}$, \ldots , $\psi_m^{(l_m)}$. Denote by $\sigma^p(H_0)$ the
following subset of $\sigma(H_0)$:
%
%
%
$$
\sigma^p(H_0):=\left\{\mu_m \in \sigma(H_0): \exists j\in\{1,\ldots,n\},\,
\exists \psi\in\HH_0(\mu_m)\,\,{\rm s.t.}\,\,\psi(q_j)\ne 0\right\}\,.
$$
%

\medskip

{\bf Proposition 1}. {\it $Q_0(z)$ is a meromorphic matrix-valued function
on the complex plain $\CC$. The set of poles of $Q_0$ is infinite and
coincides with $\sigma^p(H_0)$}.

\medskip

\noindent {\normalsize\bf Proof}. Using  Mercer's Theorem it is not hard to derive the equality
\begin{equation}
                                    \label{n3.2}
\frac{\partial [Q_0(z)]_{jk}}{\partial z}=
\sum\limits_{m=0}^{\infty}(\mu_m-z)^{-2}\sum\limits_{s=1}^{l_m}
\overline{\psi_m^{(s)}(q_j)}\psi_m^{(s)}(q_k)\,,
\end{equation}
where the series converges absolutely and locally uniformly with respect to
$z$, $z\in\CC\setminus\sigma(H_0)$. It is hence clear that $Q_0$ is
meromorphic and $\sigma^p(H_0)$ is the set of poles for $Q_0$. Suppose that
this set is finite; then there exists $m_0$ such that
$\mu_m\notin \sigma^p(H_0)$ $\forall m>m_0$. Consider the linear hull
$\cal L$ of all the eigenfunctions $\psi_m^{(s)}$, then ${\cal L}\subset
C(X)$. Fix $j\in\{1,\ldots,n\}$ and set $q=q_j$. If $\phi\in {\cal L}$,
then the relations $\bra \psi_m^{(s)}|\phi\ket=0$ $\forall m\le m_0$,
$s=1,\ldots,l_m$, imply $\delta_q(\phi):=\phi(q)=0$. Therefore the linear
functional $\delta_q$ on ${\cal L}$ is a linear combination of the linear
functionals $\bra \psi_m^{(s)}|$ $(m\le m_0$, $s=1,\ldots,l_m)$.
Since ${\cal L}$ is dense in $C(X)$ with respect to both the Hilbert and
Chebyshev norms, we conclude that $\delta_q$ is a continuous functional on
$C(X)$ with respect to the topology induced from $L^2(X)$. This
contradiction concludes the proof. \halm

\medskip

\noindent {\bf Remark 5}.
Generally speaking, the set $\sigma^p(H_0)$ depends on the tuple
$(q_1,\ldots,q_n)$ but the set  $Y=\{(q_1,\ldots q_n)\in X^n:\,
\sigma^p(H_0)=\sigma(H_0)\}$ is generic both in sense of measure
and category (i.e. the set $X^n\setminus Y$ is a zero-measure set of the
first Baire class). Moreover, if $X$ is a homogeneous manifold, then
$X=Y$ independently of the tuple $(q_1,\ldots,q_n)$.

The structure of the spectrum for an arbitrary self-adjoint extension of
the operator $S_0$ (in particular, for the point perturbation of $H_0$)
is very simple. Namely, the following proposition is an evident consequence
of theorems 14.9 and 14.10 from \cite{Nai}.

\medskip

{\bf Proposition 2}. {\it Let $\tilde H_0$ be a self-adjoint extension of
$S_0$. Then $\tilde H_0$ is bounded from below and the spectrum of
$\tilde H_0$ is purely discrete:
$\sigma(\tilde H_0)=\sigma_{\rm dis}(\tilde H_0)$}.

\medskip

The spectral properties of a Schr\"odinger operator on $\hat X$
are rather rich. Before we describe them, we settle the following notations.
{\bf For the rest of this section $H$ will denote the
Schr\"odinger operator on $\hat X$ defined by a Schr\"odinger operator
$H_0$ on $X$ and a Lagrangian subspace $\Lambda\subset\GG\oplus\GG$}.
The next theorem describes the spectral properties of $H$.

\medskip

{\bf Theorem 4}. {\it The following assertions hold.

\noindent{\rm (i)} $\sigma_{\rm ess}(H)=\sigma_{\rm ac}(H)=[0,+\infty)$;

\noindent{\rm (ii)} $\sigma_{\rm sc}(H)=\emptyset$;

\noindent{\rm (iii)} $\sigma_{\rm dis}(H)$ is a finite $($possibly, empty$)$
subset of $(-\infty,0)$;

\noindent{\rm (iv)}
$\sigma_{\rm pp}(H)\cap[0,+\infty)\subset\sigma(\tilde H_0)$, where
$\tilde H_0$ is
a self-adjoint extension of $S_0$ $($therefore, $\sigma_{\rm pp}(H)$ has
no accumulation points$)$;

\noindent{\rm (v)} the multiplicity of an eigenvalue
$E_0\in\sigma_{\rm pp}(H)$ does not exceed $2n+m$, where $m$ is the
multiplicity of $E_0$ in the spectrum of $H_0$. Moreover, let $N$ be the
number of eigenvalues $E$ of $H$ $($counting multiplicity$)$ obeying the
inequality $E<\min(0,\inf\sigma(H_0))$; then $0\le N \le 2n$.}

\medskip

\noindent {\normalsize\bf Proof}. Clearly, the spectrum of $H_{\rm N}$
possesses all the properties (i)-(v). Therefore, general theorems about
self-adjoint extensions with finite deficiency indices (Theorems 14.9 and
14.10 from \cite{Nai}, Theorem 18 from \cite{Kre}) imply properties (iii)
and (v) for the operator $H$. Furthermore, taking into account
(\ref{n2.35}) we see that the equality  $\sigma_{\rm ess}(H)=[0,+\infty)$
follows from the Weyl theorem (\cite{RSIV}, Theorem XII.14) and that the
equality $\sigma_{\rm ac}(H)=[0,+\infty)$ is a consequence of the
Birman--Kuroda theorem (\cite{RSIII}, Theorem XI.9).

Let us prove property (iv). Fix a representation of $H$ in the form
$H^{L,\,\eta}$, where $\eta\subset\{1,\ldots,2n\}$ and $L$ is a Hermitian
$2n\times 2n$-matrix. Let $E_0$, $E_0\ge 0$, be an eigenvalue of $H$ with
an eigenvector $f=(f_0,f_1,\ldots,f_n)$. For every $j=1,\ldots,n$ the
function $f_j$ belongs to $L^2(\Rs)$ and obeys the equation
$-f''_j=E_0f_j$; hence, $f_j=0$. Using the first two equations from
(\ref{n2.26}) we show that $E_0$ is an eigenvalue of
$\tilde H_0=H_0^{B,\theta}$, where $\theta=\{1,\ldots,n\}\cap\eta$ and
$B$ is related to $L$ by Eq.~(\ref{n2.28}).

It remains to prove property (ii). Denote by ${\cal L}$ the dense subspace of
all elements $f=(f_0,f_1,\ldots,f_n)$ from $\HH$ such that $f_0\in C(X)$,
$f_j\in C^{\infty}_0(0,+\infty)$, $j=1,\ldots,n$. Let ${\cal F}$ be a family
of functions which are analytic in the upper half-plane
$\CC^+=\{z\in \CC:\,{\rm Im}\,z>0\}$; we say that the family ${\cal F}$
is bounded near a point $E$, $E\in\RR$, if there exists  a neighborhood
$V$ of $E$ such that every function from ${\cal F}$ is bounded in
$V\cap\CC^+$. According to Theorem XIII.20 from \cite{RSIV} it is enough
to prove that for some countable subset $Z$ of $\RR$ the family of the
functions $z \mapsto\bra f|R(z)g\ket$, where $f$ and $g$ run through
${\cal L}$, is bounded near every point $E$, $E\in(0,+\infty)\setminus Z$.
It is clear that for $H=H_{\rm N}$ this family is bounded near the points
from $(0,+\infty)\setminus\sigma(H_0)$. Moreover, let ${\cal F}$ be the
family of functions of the form
$$
z\mapsto\int\limits_X G_0(x,q_j;z)f_0(x)\,d\lambda(x)\,,
$$
or
$$
z\mapsto \int\limits_0^{\infty}G_j(x,0;z)f_j(x)\,dx\,,
$$
where $j=1,\ldots,n$ and $f=(f_0,f_1,\ldots,f_n)\in{\cal L}$. Then the
family ${\cal F}$ is bounded near every point
from $(0,+\infty)\setminus\sigma(H_0)$. According to (\ref{n2.35})
it remains to show that there exists
a discrete  subset $Z_0\subset\RR\setminus\sigma(H_0)$ such that the
elements of the matrix $[Q(z)-\Lambda]^{-1}$
form a bounded family near every  point from $(0,+\infty)\setminus Z_0$.
Rewrite $[Q(z)-\Lambda]^{-1}$ in the form $J_\Lambda[P_\Lambda Q(z)
J_\Lambda - L]^{-1}P_\Lambda$ (see (\ref{n1.8})). The elements of the matrix
$Q(z)$ have analytic continuations from the half-plane $\CC_+$ to a
neighborhood of the set $(0,+\infty)\setminus\sigma(H_0)$; moreover,
${\rm det}\,[P_\Lambda Q(z) J_\Lambda - L]\ne 0$, if ${\rm Im}\,z>0$.
Therefore, we obtain the required property from standard
analyticity arguments. \halm

\medskip

Now we are going to define the scattering matrix for the Schr\"odinger
operator $H$ on $\hat X$ following the ideas of geometric scattering
theory (see, e.g. \cite{Mel}). First of all we note that there exists a
natural extension of $H$ to a domain of functions not belonging to
$L^2(\hat X)$. Namely, Lemma 1 defines the boundary value
operators $\Gamma^{(1)}_j$ and  $\Gamma^{(2)}_j$ for every function from
$H^2_{\rm loc}(\Rs^{(j)})$. Therefore, (\ref{n2.23}) defines
the operators $\Gamma_j$ and  $\Gamma_j$ for every function
$f=(f_0,f_1,\ldots,f_n)$ from $\DD(S_0^*)\oplus H^2_{\rm loc}(\Rs^{(1)})
\oplus\ldots\oplus H^2_{\rm loc}(\Rs^{(n)})$. Hence, we can consider the
operator $H$ to be defined on the domain, $\DD_{\rm loc}(H)$, consisting of
all functions $f$ from $\DD(S_0^*)\oplus H^2_{\rm loc}(\Rs^{(1)})
\oplus\ldots\oplus H^2_{\rm loc}(\Rs^{(n)})$ obeying the boundary condition
$(\Gamma^{(1)}f, \Gamma^{(2)}f)\in\Lambda$ (this operator takes values in
the space $L^2_{\rm loc}(\hat X)=L^2(X)\oplus L^2_{\rm loc}(\Rs^{(1)})
\oplus\ldots\oplus L^2_{\rm loc}(\Rs^{(n)})$). If $H$ is represented in the
form $H=H^{L,\,\eta}$, then the last condition may be replaced by condition
(\ref{n2.26}). To define the scattering matrix we need solutions to
the Schr\"odinger equation
\begin{equation}
                                    \label{n3.3}
Hf=k^2f\,,
\end{equation}
$f\in \DD_{\rm loc}(H)$, $k\ge 0$, the so-called {\it scattering states},
which have a special behavior in the channels $\Rs^{(j)}$.
The following theorem provides us with such solutions.

\medskip

{\bf Theorem 5} (Existence and uniqueness of scattering states).
{\it For every Schr\"odinger operator $H=H^\Lambda$ on $\hat X$ there
exists a discrete subset $Z_H$ of  $\RR$ such that the
following assertion is valid.

For a given $j\in\{1,\ldots,n\}$ and every $k>0$, $k^2\notin Z_H$, the
Schr\"odinger  equation} (\ref{n3.3}) {\it has a unique solution
$f=(f_0,f_1,\ldots,f_n)$ satisfying  the conditions}:

\medskip

\noindent(i) $f_j(x)=\exp(-ikx)+r_j(k)\exp(ikx)$,

\noindent(ii) {\it  if $l\in\{1,\ldots,n\}$ and
$l\ne j$, then $f_l(x)=t_{lj}(k)\exp(ikx)$\,,

\medskip

\noindent where $r_j(k),\,t_{lj}(k)\in\CC$}.

\medskip

\noindent {\normalsize\bf Proof}. We define $Z_H$ as the union of the
following sets: 1)~$\sigma(H_0)$; 2)~$\sigma(H_0^{B,\,\theta})$ if $H$ may
be represented in the form $H=H^{L,\,\eta}$ and $(B,\,\theta)$ is related
to $(L,\,\eta)$ with (\ref{n2.23a}), (\ref{n2.28}); 3)~the set of all
solutions to the equation ${\rm det}\,[P_\Lambda Q(E) J_\Lambda -L]=0$
where $Q(E)$ is the analytic continuation of $Q(z)$ from the upper
half-plane $\CC^+$ to $\RR^+$ (see the proof of Theorem 4). Clearly,
$Z_H$ is discrete.

Further we note that for every $\zeta=(\zeta_l)_{1\le l\le 2n}
\in\CC^{2n}=\GG$ the function $\gamma(z)\zeta=(\phi_0,\phi_1,\ldots,\phi_n)$
has the form
\begin{equation}
                                    \label{l3.6}
\phi_0(x)=\sum\limits_{m=1}^{n}\zeta_m G_0(x,q_m;z)\,,
\end{equation}
\begin{equation}
                                    \label{l3.7}
\phi_l(x)=\frac{\zeta_{l+n}}{\sqrt{-z}}\exp(-\sqrt{-z}x)\,,\quad 0<l\le n\,,
\end{equation}
(see (\ref{n2.4})), (\ref{n2.22}), and (\ref{n2.23})).
Therefore, for any $\phi\in\HH$
%
%
 %
$$                                   \label{l3.8}
\gamma^*(\bar z)\phi=(\zeta_l)_{1\le l\le 2n}\,,
$$
%
where
\begin{equation}
                                    \label{l3.9}
\zeta_l=\int\limits_X G_0(q_l,x;z)\phi_0(x)\,d\lambda(x)\,,
\quad 1\le l \le n\,;
\end{equation}
\begin{equation}
                                    \label{l3.10}
\zeta_l=\frac{1}{\sqrt{-z}}\int\limits_0^{\infty}\exp(-\sqrt{-z}x) \phi_{l-n}(x)\,dx\,,
\quad n+1\le l \le 2n\,.
\end{equation}

Fix now $k>0$, $k^2\notin Z_H$, and put $z=k^2+i\epsilon$, where
$0<\epsilon\le \epsilon_0$, with some $\epsilon_0>0$. It is clear that an
element $g$ from $\HH$ belongs to $\DD(H)$ if and only if $g=R(z)h$, where
$h$ is an element from $\HH$ (which depends on $z$). Set
$\psi=R_{\rm N}(z)h$, then $\psi\in\DD(H_{\rm N})$ and from (\ref{n2.35})
\begin{equation}
                                    \label{l3.11}
g=\psi-\gamma(z)\left[Q(z)-\Lambda\right]^{-1}\gamma^*(\bar z)(H_{\rm N}-z)
\psi\,,
\end{equation}
and
\begin{equation}
                                    \label{l3.12}
(H-z)g=(H_{\rm N}-z)\psi\,.
\end{equation}
Conversely, every function $\psi\in\DD(H_{\rm N})$ defines, by
(\ref{l3.11}), an element $g$ from $\DD(H)$ in such a way that
(\ref{l3.12}) holds. Note that according to (\ref{l3.9}) and (\ref{l3.10}),
the vector $\xi=\gamma^*(\bar z)(H_{\rm N}-z)\psi$ has the form
%
%
$$
\xi_l=\cases{\psi_0(q_l)\,, &if $1\le l \le n$\,;\cr
\noalign{\medskip}
            \psi_{l-n}(0)\,, &if $n+1\le l \le 2n$.}
$$
%
Therefore, we can rewrite (\ref{l3.11}) as
%
%
$$
g=\psi-\gamma(z)\zeta(\psi)\,,
$$
%
with
\begin{equation}
                                    \label{l3.15}
\zeta_l(\psi)=\sum\limits_{m=1}^n \left[Q(z)-\Lambda\right]_{lm}^{-1}\psi_0(q_m)+
\sum\limits_{m=n+1}^{2n} \left[Q(z)-\Lambda\right]_{lm}^{-1}\psi_{m-n}(0)\,.
\end{equation}

Now fix $j\in\{1,\ldots,n\}$ and define $\psi=(\psi_0,\psi_1,\ldots,\psi_n)$
from $L^2_{\rm loc}(\hat X)$ by
\begin{equation}
                               \label{l3.16}
\psi_l(x)=\cases{\exp(ikx)+\exp(-ikx)\,, &if $l=j$\,,\cr
\noalign{\medskip}
0\,, & otherwise\,.}
\end{equation}
It is clear that $\psi\in\DD_{\rm loc}(H_{\rm N})$ but
$\psi\notin \DD(H_{\rm N})$. To obtain a function from $\DD(H)$ we
choose for $a>0$ a cut-off function $\chi_a\in C_0^{\infty}(\Rs)$
such that $\chi_a(x)=1$ if $0 \le x \le a$, $\chi_a(x)=0$ if $x>a+1$,
and $0 \le\chi_a(x)\le 1$ $\forall x\in\Rs$. Set
$\tilde \chi_a:=(1,\chi_a,\ldots,\chi_a)$, it is clear that the product
$\tilde \chi_a\psi=(\psi_0, \chi_a\psi_1,\ldots,\chi_a\psi_n)$ is in
$\DD(H_{\rm N})$, and hence defines a function
\begin{equation}
                                    \label{l3.17}
g^{(a)}:=\tilde\chi_a\psi-\gamma(z)\zeta(\tilde\chi_a\psi)
\end{equation}
in $\DD(H)$ such that
\begin{equation}
                                    \label{l3.18}
(H-z)g^{(a)}=(H_{\rm N}-z)\tilde\chi_a\psi\,.
\end{equation}
We write the matrix $[Q(z)-\Lambda]^{-1}$ in block form,
\begin{equation}
                              \label{n3.16}
\left[Q(z)-\Lambda\right]^{-1}=\left[
\begin{array}{cc}
N(z)&W(z)\\
\noalign{\medskip}
M(z)&V(z)
\end{array}
\right]\,,
\end{equation}
where $W(z)=(w_{lm}(z))$ and $V(z)=(v_{lm}(z))$ are $n\times n$-matrices.  
From (\ref{l3.15}) we have
$$
\zeta_l(\tilde \chi_a \psi)=
\sum\limits_{m=n+1}^{2n}\left[Q(z)-\Lambda\right]^{-1}_{l,m}\psi_{m-n}(0)
$$
%
%
$$
=2\sum\limits_{m=1}^{n}\left[Q(z)-\Lambda\right]^{-1}_{l,m+n}\delta_{jm}=
2\left[Q(z)-\Lambda\right]^{-1}_{l,j+n}\,.
$$
%
In other words,
%
%
$$
\zeta_l(\tilde \chi_a\psi)=\cases{2w_{lj}(z)\,, & if $1\le l \le n$\,,\cr
\noalign{\medskip}
2v_{lj}(z)\,, & if $n+1\le l \le 2n$\,.}
$$
%
Hence, from (\ref{l3.6}), (\ref{l3.7}), and (\ref{l3.17}) we get
\begin{equation}
                              \label{l3.21}
g^{(a)}_0(x)=-2\sum\limits_{m=1}^n w_{mj}(z)G_0(x,q_m;z)\,,
\end{equation}
\begin{equation}
                              \label{l3.22}
g^{(a)}_l(x)=\delta_{lj}\chi_a(x)\psi_j(x)-\frac{2v_{lj}(z)}{\sqrt{-z}}
\exp(-\sqrt{-z}x)\,,\quad 0< l \le n\,.
\end{equation}

Passing to the limit $a\to\infty$ in (\ref{l3.21}) and (\ref{l3.22}),
we obtain $g=(g_0,g_1,\ldots,g_n)\in L^2_{\rm loc}(\hat X)$ with
%
%
$$
g_0(x)=-2\sum\limits_{m=1}^n w_{mj}(z)G_0(x,q_m;z)\,,
$$
%
%
%
$$
g_l(x)=\delta_{lj}\psi_j(x)-\frac{2v_{lj}(z)}{\sqrt{-z}}
\exp(-\sqrt{-z}x)\,,\quad 0< l \le n\,.
$$
%
Moreover, since $\Gamma^{(1)}g^{(a)}$ and $\Gamma^{(2)}g^{(a)}$ are
independent of $a$, $g$ satisfies the boundary condition $(\Gamma^{(1)}g,
\Gamma^{(2)}g)\in\Lambda$, and from (\ref{l3.18}) we have
%
%
%
$$
(H-z)g=(H_{\rm N}-z)\psi\,.
$$
%
In the limit $\epsilon \to 0$ we have $\sqrt{-z}\to -ik$, whereas $g$ has
a limit  $f$ in $L^2_{\rm loc}(\hat X)$ such that
%
%
\begin{eqnarray}
                              \label{l3.27}
f_0(x)&=&-2\sum\limits_{m=1}^n w_{mj}(k^2)G_0(x,q_m;k^2)\,,\nonumber\\
%
%
f_l(x)&=&\delta_{lj}\psi_j(x)+\frac{2v_{lj}(k^2)}{ik}
\exp(ikx)\,,\quad 0< l \le n\,,
\end{eqnarray}
since $k^2\notin \sigma(H_0)$. Moreover, in the sense of distributions,
%
%
$$
(H-z)g\to (H-k^2)f\,,
$$
%
%
$$
(H_{\rm N}-z)\psi\to (H_{\rm N}-k^2)\psi=0\,.
$$
%
Hence, $(H-k^2)f=0$. Further, $f$ satisfies the boundary condition
$(\Gamma^{(1)}f,\Gamma^{(2)}f)\in\Lambda$. Indeed, since
$k^2\notin \sigma(H_0)$ we have
$$
a_m(G_0(\cdot,q_l;z))\to \delta_{lm}\,,
\quad b_m(G_0(\cdot,q_l;z))\to Q^{ml}_0(k^2)
$$
as $z\to k^2$ (see ({\ref{n2.8}) and (\ref{n2.10})). On the other hand,
direct calculations show that
$\Gamma_l^{(1)}g_l\to \Gamma_l^{(1)}f_l$ and
$\Gamma_l^{(2)}g_l\to \Gamma_l^{(2)}f_l$
as $z\to k^2$ ($l=1,\ldots,n;$).
Finally, from (\ref{l3.27}) we get the properties (i) and (ii) with
\begin{eqnarray}
                              \label{n3.25}
r_j(k)&=&1-2ik^{-1}v_{jj}(k^2)\,,\nonumber\\
t_{lj}(k)&=&-2ik^{-1}v_{lj}(k^2)\,.
\end{eqnarray}

The proof is completed by establishing the uniqueness of $f$ which follows
from Lemma 7 below. \halm

\medskip

{\bf Lemma 7}. {\it Let $f$ be a solution to the
Schr\"odinger equation} (\ref{n3.3}) {\it for some $k^2\notin Z_H$, with
the property that for all $l$, $1\le l\le n$, }
\begin{equation}
                              \label{l3.30}
f_l(x)=\alpha_l\exp(ikx)\,,
\end{equation}
{\it for some $\alpha_l\in\CC$. Then $f=0$}.

\medskip

\noindent {\normalsize\bf Proof}. Take $\zep=k^2+i\epsilon$ and $a>0$ as
in the proof of the theorem. It is evident that $\tilde\chi_a f\in\DD(H)$;
denote
%
%
$$
\psi^{(\epsilon)}=R_{\rm N}(\zep)(H-\zep)\tilde\chi_a f\,.
$$
%
Then
\begin{equation}
                             \label{l3.32}
(H_{\rm N}-\zep)\psi^{(\epsilon)}=(H-\zep)\tilde\chi_a f\,,
\end{equation}
and
\begin{equation}
                             \label{l3.33}
\tilde\chi_a f=\psi^{(\epsilon)}-\gamma(\zep)\zeta(\psi^{(\epsilon)})\,.
\end{equation}
Consider the integral kernel $G^{\rm N}_l(x,y;z)$ of the operator $R^{\rm N}_l(z)=
(H^{\rm N}_l-z)^{-1}$, then
%
%
$$
G^{\rm N}_l(x,y;\zep)\to \frac{i}{2k}\left[\exp(ik|x-y|)+\exp(ik(x+y))\right]
\equiv G^{\rm N}_l(x,y;k^2)
$$
%
as $\epsilon \to 0$ (see (\ref{n2.5})). Denote by $\HH_l^{(a)}$ the
following subspace of $\HH_l=L^2(\Rs^{(l)})$:
%
%
$$
\HH_l^{(a)}:=\{\phi\in L^2(\Rs^{(l)}):\,{\rm supp}\,\phi\subset[0,a+1]\}\,;
$$
%
then $G_l^{\rm N}(x,y;z)$ is the kernel of a continuous linear operator from
$\HH_l^{(a)}$ to $L^2_{\rm loc}(\Rs^{(l)})$ (recall that $L^2_{\rm loc}(\Rs^{(l)})$ is endowed
with the topology of $L^2$-convergence on compact subsets of $\Rs^{(l)}$).
Set $\HH^{(a)}:=\HH_0\oplus\HH_1^{(a)}\oplus\ldots\oplus\HH_n^{(a)}$, then
$$
R_{\rm N}(k^2):=
R_0(k^2)\oplus R^{\rm N}_1(k^2)\oplus\ldots\oplus R^{\rm N}_n(k^2)
$$
is a continuous linear operator from $\HH^{(a)}$ to $L^2_{\rm loc}(\hat X)$.
Moreover, if $h\in\HH^{(a)}$ then $R_{\rm N}(\zep)h\to R_{\rm N}(k^2)h$ as
$\epsilon \to 0$; in particular $\psi^{(\epsilon)}\to R_{\rm N}(k^2)(H-k^2)
\tilde\chi_a f =: \psi^{(0)}$. Hence, $\tilde\chi_a \psi^{(\epsilon)}\to
\tilde\chi_a \psi^{(0)}$ in $L^2(\hat X)$ as $\epsilon \to 0$, too.
Fix $\epsilon_0>0$ and put $z_0=k^2+i\epsilon_0$, then
$$
(H_{\rm N}-z_0)\psi^{(\epsilon)}=(H_{\rm N}-\zep)\psi^{(\epsilon)}+
(\zep-z_0)\psi^{(\epsilon)}=
$$
%
%
$$
(H-\zep)\tilde\chi_a f+ (\zep-z_0)\psi^{(\epsilon)}\,.
$$
Therefore,
$(H_{\rm N}-z_0)\psi^{(\epsilon)}$ has a limit in $L^2_{\rm loc}(\hat X)$
as $\epsilon \to 0$. Consequently,
$$
\psi_l^{(\epsilon)}(x)\to \psi_l^{(0)}(x)\,,
\qquad \frac{d}{dx}\psi_l^{(\epsilon)}(x)\to\frac{d}{dx}\psi_l^{(0)}
$$
locally uniformly on $\Rs^{(l)}$ for each $l$, $1\le l \le n$. Now we have
$$
(H_{\rm N}-z_0)\tilde \chi_a\psi^{(\epsilon)}=
$$
\begin{equation}
                            \label{l3.37}
\tilde \chi_a (H_{\rm N}-z_0)\psi^{(\epsilon)}
-2\left(0,\chi'_a(\psi^{(\epsilon)}_1)',\ldots,\chi'_a(\psi^{(\epsilon)}_n)'
\right)-\left(0,\chi''_a\psi^{(\epsilon)}_1,\ldots,
\chi''_a\psi^{(\epsilon)}_n\right)\,.
\end{equation}
It follows from (\ref{l3.37}) that also
$(H_{\rm N}-z_0)\tilde \chi_a\psi^{(\epsilon)}$ has a limit in
$L^2(\hat X)$, and therefore, $\tilde \chi_a\psi^{(\epsilon)}$ has a limit
in the graph topology of $\DD(H_{\rm N})$ as $\epsilon\to 0$. Thus,
$\tilde \chi_a\psi^{(0)}\in\DD(H_{\rm N})$ and
$(H_{\rm N}-z_0)\tilde \chi_a\psi^{(\epsilon)}\to (H_{\rm N}-z_0)\tilde
\chi_a\psi^{(0)}$ in $L^2(\hat X)$.

Now (\ref{l3.32}) implies that
$$
(H_{\rm N}-\zep)\tilde \chi_a\psi^{(\epsilon)}=
$$
\begin{equation}
                            \label{l3.38}
\tilde \chi_a (H-\zep)\tilde \chi_a f-
2\left(0,\chi'_a(\psi^{(\epsilon)}_1)',
\ldots,\chi'_a(\psi^{(\epsilon)}_n)'\right)-
\left(0,\chi''_a\psi^{(\epsilon)}_1,\ldots,\chi''_a
\psi^{(\epsilon)}_n\right)\,.
\end{equation}
Since $\left(\tilde \chi_a(H-k^2)\tilde\chi_a f\right)(x)=0$ if
$x\in\Rs^{(l)}$ and $0\le x \le a$, we get from (\ref{l3.38}) by passing
to the limit $\epsilon\to 0$
%
%
$$
(H_{\rm N}-\lambda)\hat\chi_a\psi^{(0)}(x)=0\,,\quad {\rm if}\,\,\,
x\in\Rs^{(l)}\,,\,\,\,\,0\le x \le a\,.
$$
%
Because $\tilde\chi_a\psi^{(0)}\in\DD(H_{\rm N})$, we have for every
$l\ge 1$:
%
%
$$
\psi_l^{(0)}(x)=c_l\left(\exp(ikx)+\exp(-ikx)\right)\,,
\quad{\rm if}\,\,\, x\in[0,a]\,.
$$
Now, we turn to (\ref{l3.33}); for $l\ge 1$ this equality reads
\begin{equation}
                               \label{l3.41}
\chi_a(x)f_l(x)=
\psi_l^{(\epsilon)}(x)-\frac{\zeta_{l+n}(\psi^{(\epsilon)})}{\sqrt{-z}}
\exp(-\sqrt{-z}x)\,,
\end{equation}
where
%
%
$$
\zeta_{l+n}(\psi^{(\epsilon)})=\sum\limits_{m=1}^n \left[Q(z)-
\Lambda\right]_{l+n,m}^{-1}\psi_0^{(\epsilon)}(q_m)+
\sum\limits_{m=n+1}^{2n} \left[Q(z)-\Lambda\right]_{l+n,m}^{-1}\psi_{m-n}^
{(\epsilon)}(0)
$$
%
(see (\ref{l3.7}) and (\ref{l3.15})). By definition of $\psi^{(\epsilon)}$
we have $\psi^{(\epsilon)}_0=R_0(k^2+i\epsilon)(S_0^*-k^2-i\epsilon)f_0$.
Moreover
$R_0(z)$ is a continuous mapping from $L^2(X)$ to $\DD(H_0)$ endowed with
the graph topology and hence, a continuous mapping from $L^2(X)$ to $C(X)$
which continuously depends on $\epsilon\in[0,\epsilon_0]$. Therefore,
$\psi^{(\epsilon)}_0(q_m)\to \psi^{(0)}_0(q_m)$ as $\epsilon\to 0$.
Hence, we obtain from (\ref{l3.41}) that for $l\ge 1$
\begin{equation}
                               \label{l3.43}
\chi_a(x)f_l(x)=\psi_l^{(0)}(x)+\frac{\zeta_{l+n}(\psi^{(0)})}{ik}
\exp(ikx)\,,
\end{equation}
where the coefficients
\begin{equation}
                                    \label{l3.44}
\zeta_{l+n}(\psi^{(0)})=\sum\limits_{m=1}^n \left[Q(k^2)-
\Lambda\right]_{l+n,m}^{-1}\psi_0^{(0)}(q_m)+
\sum\limits_{m=n+1}^{2n} \left[Q(k^2)-\Lambda\right]_{l+n,m}^{-1}
\psi_{m-n}^{(0)}(0)
\end{equation}
are well defined because $k^2\notin Z_H$. Moreover,
$\psi_0^{(0)}(q)=0$, since $k^2\notin \sigma(H_0)$,
and (\ref{l3.43}) implies that for $x\in[0,a]$ and $l \ge 1$ the functions
$f_l$ have the form
$$
f_l(x)=c_l\exp(-ikx)+c'_l\exp(ikx)\,.
$$
Comparing with (\ref{l3.30}), we get $c_l=0$ and hence,
$\psi_l^{(0)}(x)=0$ for $x\in[0,a]$. Returning to (\ref{l3.43})
and (\ref{l3.44}) we obtain that $f_l(x)=0$ for
$x\in[0,a]$. Since $a$ is arbitrary, $f_l=0$ $\forall l=1,\ldots,n$.

Using (\ref{n2.26}) we see that $f_0$ satisfies the boundary conditions
(\ref{n2.27}). Moreover, by the hypothesis of the lemma, $f_0$ is a solution
to the equation $(S_0^*-k^2)f_0=0$. Since
$k^2\notin \sigma(H_0^{B,\,\theta})$, we get $f_0=0$. Thus, the lemma
is proven, and the proof of Theorem 5 is completed. \halm

\medskip

Property (i) of Theorem 5 means that the function $f_j(x)$ represents a
superposition of an incoming wave $\exp(-ikx)$ and a reflected wave
$r_j(k)\exp(ikx)$ in the channel $\Rs^{(j)}$.

\medskip

\noindent {\bf Definition}. $r_j(k)$ is called
{\it the reflection amplitude} for $H$ in the channel
$\Rs^{(j))}$ at energy $E=k^2$. The quantity $R_j(k)=|r_j(k)|^2$ is called
{\it the reflection coefficient} ({\it or the reflection
probability}) in the channel $\Rs^{(j)}$.

\medskip

Condition (ii) in Theorem 5 means that the function $f_l(x)$ ($l\ne j$)
represents an outgoing wave $t_{lj}(k)\exp(ikx)$ in the channel
$\Rs^{(l)}$.

\medskip

\noindent {\bf Definition}. $t_{lj}(k)$ is called {\it the transmission
amplitude} for $H$ from the channel $\Rs^{(j)}$ to the channel $\Rs^{(l)}$
at energy $E=k^2$. The quantity $T_{lj}(k)=|t_{lj}(k)|^2$ is called
{\it the transmission coefficient} ({\it or the transmission
probability}) from $\Rs^{(j)}$ to $\Rs^{(l)}$.

\medskip

Set
%
%
%
$$
s_{lj}(k)=\cases{r_j(k)\,,& if $l=j$\,;\cr
\noalign{\medskip}
                 t_{lj}(k)\,,& otherwise\,.}
$$

\medskip

The matrix $\Sigma(k)=\left(s_{lj}(k)\right)_{1\le l,j\le n}$ is called the
{\it scattering matrix} for $H$. We stress that $\Sigma(k)$ is defined for
all $k>0$ with the exception of a discrete subset
$Z_H$ of $\RR$.

\medskip

{\bf Theorem 6}. {\it The scattering matrix $\Sigma(k)$ is unitary for all
$k\in Z_H$. If the matrix $[Q(k^2)-\Lambda]^{-1}$ is represented in the form}
(\ref{n3.16}), {\it then}
\begin{equation}
                                \label{n3.34}
\Sigma(k)=I-2ik^{-1}V(k^2)\,.
\end{equation}
{\it If $\Lambda$ is the graph of a Hermitian operator $L$ in $\GG$
and the matrix $L$ is represented in the form}  (\ref{n2.28}), {\it then}
\begin{equation}
                               \label{n3.35}
\Sigma(k)=\left[C+A^*(Q_0(k^2)-B)^{-1}A+ik^{-1}I\right]
\left[C+A^*(Q_0(k^2)-B)^{-1}A-ik^{-1}I\right]^{-1}\,.
\end{equation}
{\it In particular, if the matrix $A$ is invertible, and $C$ is a scalar
matrix $($i.e. $C=\gamma I$, $\gamma \in \RR)$, then}
$$
\Sigma(k)=\left[ikI+(ik\gamma-1)A^{-1}(Q_0(k^2)-B)A^{*-1}\right]
\left[ikI+(ik\gamma+1)A^{-1}(Q_0(k^2)-B)A^{*-1}\right]^{-1}=
$$
\begin{equation}
                               \label{n3.36}
A^{-1}\left[ikAA^*+(ik\gamma-1)(Q_0(k^2)-B)\right]
\left[ikAA^*+(ik\gamma+1)(Q_0(k^2)-B)\right]^{-1}A\,.
\end{equation}

\medskip

{\normalsize\bf Proof}. (\ref{n3.34}) follows immediately from (\ref{n3.25}) as
obtained in the proof of Theorem 5. To get (\ref{n3.35}) we use
the Frobenius formula for the inverse of a block-matrix \cite{HJ}:
$$
\left[
\begin{tabular}{p{5mm}p{5mm}}
$A_{11}$&$A_{12}$\cr
\noalign{\medskip}
$A_{21}$&$A_{22}$\cr
\end{tabular}
\right]^{-1}=
$$
\begin{equation}
                                        \label{n3.37}
\left[
\begin{tabular}{p{51mm}p{51mm}}
$[A_{11}-A_{12}A_{22}^{-1}A_{21}]^{-1}$&
$A_{11}^{-1}A_{12}[A_{21}A_{11}^{-1}A_{12}-A_{22}]^{-1}$\cr
\noalign{\bigskip}
$[A_{21}A_{11}^{-1}A_{12}-A_{22}]^{-1}A_{21}A_{11}^{-1}$&
$[A_{22}-A_{21}A_{11}^{-1}A_{12}]^{-1}$\cr
\end{tabular}
\right]_{{}^.}
\end{equation}
Since
%
%
$$
Q(z)-L=
\left[
\begin{tabular}{p{20mm}p{20mm}}
$Q_0(z)-B$&\quad$-A$\cr
\noalign{\bigskip}
$\quad-A^*$&$ik^{-1}I-C$\cr
\end{tabular}
\right]
$$
%
then
%
%
$$
V(k^2)=\left[ik^{-1}I-C-A^*(Q_0(k^2)-B)^{-1}A\right]^{-1}\,,
$$
%
and hence
\begin{equation}
                                \label{n3.40}
\Sigma(k)=I-2\left[I+ik(C+A^*(Q_0(k^2)-B)^{-1}A)\right]^{-1}.
\end{equation}
Now we get (\ref{n3.35}) from (\ref{n3.40}) after some elementary algebra,
(\ref{n3.36}) is an evident consequence of (\ref{n3.35}).

In particular, (\ref{n3.35}) shows that $\Sigma(k)$ is the Cayley transform
of the Hermitian matrix $C+A^*(Q_0(k^2)-B)^{-1}A$. Hence, in case
$\Lambda$ is the graph of an Hermitian operator in $\GG$, $\Sigma(k)$ is a
unitary matrix. To prove the general case we use Proposition E.  \halm

\bigskip

In the notation of Example (2) from Section 2, (\ref{n3.35}) gives
the scattering matrix for the Schr\"odinger operator of Dirichlet type,
$H^L_{\rm D}$. We now derive an explicit expression for the scattering
matrix in case of an arbitrary Schr\"odinger operator $H^{\Lambda}$.
It is convenient to write the boundary conditions in the form (\ref{n1.4a})
where the $(2n\times 2n$)-matrices $L$ and $M$ have block structure
\begin{equation}
                                    \label{n3.40a}
L=\left[
\begin{tabular}{p{5mm}p{5mm}}
$B$&$A_1$\cr
\noalign{\medskip}
$A_2$&$C$\cr
\end{tabular}
\right]\,,\quad
M=\left[
\begin{tabular}{p{5mm}p{5mm}}
$Y$&$X_1$\cr
\noalign{\medskip}
$X_2$&$Z$\cr
\end{tabular}
\right]\,{}_{,}
\end{equation}
and satisfy conditions (a) and (b) from Proposition B. In particular, condition (a) is equivalent to
the  relations
$$
BY^*+A_1X_1^*=YB^*+X_1A_1^*\,,
$$
$$
BX_2^*+A_1Z^*=YA_2^*+X_1C^*\,,
$$
\begin{equation}
                                       \label{n3.40b}
A_2X_2^*+CZ^*=X_2A_2^*+ZC^*\,.
\end{equation}

Suppose for the moment that $M$ is invertible, then condition (\ref{n1.4a})
reads: $\Gamma^{(2)}x=\Lambda\Gamma^{(1)}x$ where $\Lambda=M^{-1}L$.
Therefore, using the Frobenius formula (\ref{n3.37}) we obtain the following
expression for the matrix $V$ from (\ref{n3.16}):
$$
V=\left[ik^{-1}Z-C-(X_2Q_0-A_2)(YQ_0-B)^{-1}(ik^{-1}X_1-A_1)\right]^{-1}
$$
\begin{equation}
                                   \label{n3.40c}
\cdot\,\left[Z-(X_2Q_0-A_2)(YQ_0-B)^{-1}X_1\right]\,.
\end{equation}
Substituting (\ref{n3.40c}) in (\ref{n3.34}) we finally obtain
$$
\Sigma(k)=
\left[ikC+Z-(X_2Q_0(k^2)-A_2)(YQ_0(k^2)-B)^{-1}(ikA_1+X_1)\right]^{-1}
$$
\begin{equation}
                                         \label{n3.40d}
\cdot\,\left[ikC-Z-(X_2Q_0(k^2)-A_2)(YQ_0(k^2)-B)^{-1}(ikA_1-X_1)\right]\,.
\end{equation}
Since invertible matrices are dense in the space of all
$(2n\times 2n)$-matrices, expression (\ref{n3.40d}) is valid for all
boundary conditions of the form (\ref{n1.4a}). In particular, if $M=I$
then we recover (\ref{n3.35}).

Next we consider some particular cases of (\ref{n3.40d}) and determine the
scattering matrices for the Schr\"odinger operators from Examples (1), (3),
and (4) of Section~2 (Example (2) contains operators of Dirichlet type
which are covered by (\ref{n3.35})).

\medskip

\noindent 1. Consider Example (1) from Section 2. Using the notations there
we have
%
%
$$
L=\left[
\begin{tabular}{p{5mm}p{5mm}}
$I$&$0$\cr
\noalign{\medskip}
$0$&$I$\cr
\end{tabular}
\right]\,,\quad
M=\left[
\begin{tabular}{p{5mm}p{5mm}}
$-B$&$\,\,\,A$\cr
\noalign{\medskip}
$\,\,A^*$&$-C$\cr
\end{tabular}
\right]\,{}_{.}
%
$$
Therefore,
\begin{equation}
                                            \label{n3.40f}
\Sigma(k)=\left[ik+C-A^*(Q_0^{-1}(k^2)+B)^{-1}A\right]
\left[ik-C+A^*(Q_0^{-1}(k^2)+B)^{-1}A\right]^{-1}\,.
\end{equation}

\medskip

\noindent 2. Now we turn to Example (3) from Section 2. In this case
%
%
$$
L=\left[
\begin{tabular}{p{5mm}p{5mm}}
$-I$&$A$\cr
\noalign{\medskip}
$\,\,0$&$C$\cr
\end{tabular}
\right]\,,\quad
M=\left[
\begin{tabular}{p{5mm}p{5mm}}
$B$&$0$\cr
\noalign{\medskip}
$A^*$&$I$\cr
\end{tabular}
\right]\,{}_{.}
%
$$
Hence
\begin{equation}
                                            \label{n3.40h}
\Sigma(k)=\left[C+ik^{-1}-A^*(Q_0^{-1}(k^2)+B)^{-1}A\right]
\left[C-ik^{-1}-A^*(Q_0^{-1}(k^2)+B)^{-1}A\right]^{-1}\,.
\end{equation}

\medskip

\noindent 3. Finally, let us consider an operator of Neumann type
$H^L_{\rm N}$ (as in Example (4) of Section 2). Now
%
%
$$
L=\left[
\begin{tabular}{p{5mm}p{5mm}}
$\,\,\,B$&$\,\,\,\,0$\cr
\noalign{\medskip}
$-A^*$&$-I$\cr
\end{tabular}
\right]\,,\quad
M=\left[
\begin{tabular}{p{5mm}p{5mm}}
$I$&$-A$\cr
\noalign{\medskip}
$0$&$\,\,\,C$\cr
\end{tabular}
\right]\,{}_{,}
%
$$
and we get a simple expression for $\Sigma(k)$, which is
similar to (\ref{n3.35}):
\begin{equation}
                               \label{n3.44}
\Sigma(k)=\left[ikI+C+A^*(Q_0(k^2)-B)^{-1}A\right]
\left[ikI-C-A^*(Q_0(k^2)-B)^{-1}A\right]^{-1}\,.
\end{equation}
In particular, if the matrix $A$ is invertible and $C$ is a Hermitian scalar
matrix ($C=\gamma I$), then
$$
\Sigma(k)=\left[(ik+\gamma)A^{-1}(Q_0(k^2)-B)A^{*-1}+I\right]
\left[(ik-\gamma)A^{-1}(Q_0(k^2)-B)A^{*-1}-I\right]^{-1}=
$$
\begin{equation}
                               \label{n3.45}
A^{-1}\left[(ik+\gamma)(Q_0(k^2)-B)+AA^*\right]
\left[(ik-\gamma)(Q_0(k^2)-B)-AA^*\right]^{-1}A\,.
\end{equation}

\medskip

\noindent {\bf Remark 6}. There is another way to get (\ref{n3.44}) which is
similar to the derivation of (\ref{n3.35}). Namely, if we use (\ref{n2.36})
to express the resolvent of $H^L_{\rm N}$ and start with the function
%
%
$$
\psi_j(x)=\exp(ikx)-\exp(-ikx)
$$
%
in the channel $\Rs^{(j)}$ (instead of the function (\ref{l3.16})), then we
get, arguing as in the proof of Theorem 5,
\begin{equation}
                             \label{n3.42}
\Sigma(k)=2ikV_{\rm D}(k^2)-1,
\end{equation}
where $V_{\rm D}$ is the $n\times n$-matrix in the block representation of
$[Q_{\rm D}(z)-\Lambda]^{-1}$:
\begin{equation}
                            \label{n3.43}
\left[Q_{\rm D}(z)-\Lambda\right]^{-1}=\left[
\begin{array}{cc}
N_{\rm D}(z)&W_{\rm D}(z)\\
\noalign{\medskip}
M_{\rm D}(z)&V_{\rm D}(z)
\end{array}
\right]\,.
\end{equation}
(Note that $Q_{\rm D}$ and $\gamma_{\rm D}$ in (\ref{n2.36}) are given by
(\ref{n2.6}), (\ref{n2.10}), (\ref{n2.23b})). From (\ref{n3.42}) and
(\ref{n3.43}) we get (\ref{n3.44}) again.

\medskip

It is interesting to note that for the Schr\"odinger operator
$H=H_0^{B,\,\theta}\oplus H_1^{\rm N}\oplus\ldots\oplus H_n^{\rm N}$
(see Example 1 from Section 2), we have $\Sigma(k)=I$ independently of $k$,
i.e. we have in each channel a complete reflection without phase shift.
On the other hand, if
$H=H_0^{B,\,\theta}\oplus H_1^{\rm D}\oplus\ldots\oplus H_n^{\rm D}$ (see
Example 3 from Section 2), then  $\Sigma(k)=-I$ independently of $k$, i.e.
there is complete reflection in each channel with a phase shift of
magnitude~$\pi$.

\medskip

\noindent {\bf Remark 7}. With obvious modifications, the results of this
section are valid for the case $d=0$ ($X$ is a finite set of isolated
points). This case is not empty as it seems at first sight. For example,
if we put in (\ref{n3.40d}) $A_1=A_2=X_1=X_2=0$, then
\begin{equation}
                                 \label{n3.45a}
\Sigma(k)=[ikC+Z]^{-1}[ikC-Z]\,.
\end{equation}
This is the scattering matrix for a system of quantum wires with a
single common vertex derived in \cite{KS}.

\vskip 5ex

\noindent{\bf 4.~The cases of one and two "horns"}

\bigskip

\noindent We consider now the most interesting particular cases.
For $n=1$ we denote $q_1$ as $q$ and $r_1$ as $r$ for simplicity.
If $H$ is an operator of Dirichlet type, i.e. if $H=H^L_{\rm D}$ (see
Example 2 of Section 2), then
%
%
$$
Q(k^2)-L=\left[\matrix{Q_0(k^2)-\beta&\alpha\cr
\bar\alpha&ik^{-1}-\gamma\cr}\right],
$$
%
where $\alpha\in\CC$, $\beta, \gamma\in\RR$ are arbitrary. In this case
$\Sigma(k)$ coincides with the reflection amplitude $r(k)$. Using
(\ref{n3.36}), we get
\begin{equation}
                 \label{4.2}
\Sigma(k)=\frac{(i\gamma k-1)(Q_0(k^2)-\beta)+i|\alpha|^2k}
{(i\gamma k+1)(Q_0(k^2)-\beta)+i|\alpha|^2k}\,.
\end{equation}
Obviously, we have $R(k)\equiv 1$ for the reflection coefficient.

Similarly, for the operator of Neumann type
$H=H^L_{\rm N}$ (see Example 4 of Section 2) we obtain
\begin{equation}
                            \label{4.3}
\Sigma(k)=\frac{(ik+\gamma)(Q_0(k^2)-\beta)+|\alpha|^2}
{(ik-\gamma)(Q_0(k^2)-\beta)-|\alpha|^2}\,.
\end{equation}
It is convenient to write
%
%
$$
\Sigma(k)=:e^{i\Phi(k)}\,,
$$
%
where $\Phi(k)$ is the so-called {\it scattering phase}.

\medskip

Equations (\ref{4.2}) and (\ref{4.3}) have interesting consequences.
First we recall that a point $E\in\RR$ is called a {\it point level}
of the operator $\tilde H_0=H_0^B\equiv H_0^\beta$, if $Q(E)-\beta=0$.
The spectrum of $\tilde H_0$ (recall that $\tilde H_0$ is a point
perturbation of $H_0$) consists of all point levels and all multiple
eigenvalues of the unperturbed operator  $H_0$.

\medskip

{\bf Theorem 7}. 1. {\it Let $n=1$ and  $H$ be a Schr\"odinger operator of
Dirichlet type:
$H=H^L_{\rm D}$. Then the following assertions hold.}

(1a) $\Sigma(k)=1$ ({\it i.e.} $\Phi(k)\equiv 0$} mod\,$2\pi$) {\it if and only if $k^2$ is an energy
level for the point perturbation $H_0^\beta$ of $H_0$}\,.

(1b) {\it Let, in addition, $\gamma=0$. Then  $\Sigma(k)=-1$
$($i.e. $\Phi(k)\equiv \pi$ ${\rm mod}\, 2\pi)$ if and only if
$k^2\in\sigma^p(H_0)$.
Therefore, for a generic point $q\in X$, $\Sigma(k)=-1$
if and only if $k^2\in\sigma(H_0)$.}

2. {\it Let $n=1$ and  $H$ be a Schr\"odinger operator of Neumann type:
$H=H^L_{\rm N}$. Then the following assertions hold.}

(2a) $\Sigma(k)=-1$ ({\it i.e.} $\Phi(k)\equiv \pi $ mod\,$2\pi$)
{\it if and only if $k^2$ is an energy
level for the point perturbation $H_0^\beta$ of $H_0$}\,.

(2b) {\it Let, in addition, $\gamma=0$. Then  $\Sigma(k)=1$
$($i.e. $\Phi(k)\equiv 0$ ${\rm mod}\, 2\pi)$ if and only if
$k^2\in\sigma^p(H_0)$.
Therefore, for a generic point $q\in X$, $\Sigma(k)=1$
if and only if $k^2\in\sigma(H_0)$.}

\medskip

{\normalsize\bf Proof}. The theorem is an immediate consequence
of (\ref{4.2}) and (\ref{4.3}). \halm

\medskip

Theorem 7 shows that by means of an infinitely thin  horn $\Rs$ attached
to the manifold $X$ at a point $q$ we can "hear" the positive point levels
of a point perturbation of $H_0$ at the point $q$. Moreover, if $q$ is a
generic point, we can hear the positive part of the spectrum of the
Schr\"odinger operator $H_0$ on $X$. Therefore, we can think of the horn
$\Rs$ {\it as a kind of quantum stethoscope}.

Next we consider the case of two horns ($n=2$) in some detail. For
simplicity we shall write
$$
\tQ(k^2)=Q_0(k^2)-B,
$$
%
where $B$ is a given Hermitian $2\times2$-matrix. We start with the
Schr\"odinger operator $H$ of Dirichlet type, $H=H_{\rm D}^L$.
Let $A=(\alpha_{jl})$ be an invertible $2\times 2$-matrix, $C=\gamma I$
($\gamma\in\RR$) a scalar $2\times 2$-matrix. We shall denote the matrix
$AA^*$ by $N$:
$$
N\equiv\left[
\begin{array}{cc}
\nu_{11}&\nu_{12}\cr
\nu_{21}&\nu_{22}\cr
\end{array}
\right]=
\left[
\begin{array}{cc}
|\alpha_{11}|^2+|\alpha_{12}|^2& \alpha_{11}\bar\alpha_{21}+
\alpha_{12}\bar\alpha_{22}\cr
\alpha_{21}\bar\alpha_{11}+\alpha_{22}
\bar\alpha_{12}&|\alpha_{22}|^2+|\alpha_{21}|^2\cr
\end{array}
\right]\,{}_{.}
$$
%
Further we set
$$
\Delta(k)=
(k^2\gamma-ik)\left(\nu_{12}\tQ_{21}(k^2)+\nu_{21}\tQ_{12}(k^2)-
\nu_{11}\tQ_{22}(k^2)-\nu_{22}\tQ_{11}(k^2)\right)+
$$
$$
(ik\gamma+1)^2\det\tQ(k^2) - k^2|\det A|^2\,,
$$

$$
M_{11}(k)=
(k^2\gamma+ik)\left(\nu_{21}\tQ_{12}(k^2)-\nu_{22}\tQ_{11}(k^2)\right)+
$$
$$
(k^2\gamma-ik)\left(\nu_{12}\tQ_{21}(k^2)-\nu_{11}\tQ_{22}(k^2)\right)
-(k^2\gamma^2+1)\det\tQ(k^2)-k^2|\det A|^2\,,
$$
%

$$
M_{22}(k)=
(k^2\gamma+ik)\left(\nu_{12}\tQ_{21}(k^2)-\nu_{11}\tQ_{22}(k^2)\right)+
$$
$$
(k^2\gamma-ik)\left(\nu_{21}\tQ_{12}(k^2)-\nu_{22}\tQ_{11}(k^2)\right)
-(k^2\gamma^2+1)\det\tQ(k^2)-k^2|\det A|^2\,,
$$
%

$$
M_{12}(k)=2ik\left(\nu_{12}\tQ_{11}(k^2)-\nu_{11}\tQ_{12}(k^2)\right)\,,
$$
%

$$
M_{21}(k)=2ik\left(\nu_{21}\tQ_{22}(k^2)-\nu_{22}\tQ_{21}(k^2)\right)\,.
$$
%

\noindent Then we have for the elements of the scattering matrix $\Sigma(k)$:
\begin{equation}
                                      \label{a8}
s_{11}(k)=\frac{\alpha_{11}\alpha_{22}M_{11}(k)-
\alpha_{12}\alpha_{21}M_{22}(k)+\alpha_{21}\alpha_{22}M_{12}(k)-
\alpha_{11}\alpha_{12}M_{21}(k)}{\det A \,\,\,\Delta(k)}\,,
\end{equation}
\begin{equation}
                                      \label{a9}
s_{22}(k)=\frac{\alpha_{11}\alpha_{22}M_{22}(k)-
\alpha_{12}\alpha_{21}M_{11}(k)+\alpha_{11}\alpha_{12}M_{21}(k)-
\alpha_{22}\alpha_{21}M_{12}(k)}{\det A \,\,\,\Delta(k)}\,,
\end{equation}
\begin{equation}
                                      \label{a10}
s_{12}(k)=\frac{\alpha_{12}\alpha_{22}\left(M_{11}(k)-M_{22}(k)\right)
+\alpha_{22}^2M_{12}(k)
-\alpha_{12}^2M_{21}(k)}{\det A \,\,\,\Delta(k)}\,,
\end{equation}
\begin{equation}
                                      \label{a11}
s_{21}(k)=\frac{\alpha_{11}\alpha_{21}\left(M_{22}(k)-M_{11}(k)\right)
+\alpha_{11}^2M_{21}(k)
-\alpha_{21}^2M_{12}(k)}{\det A \,\,\,\Delta(k)}\,.
\end{equation}

Shortly, we have 
\begin{equation}
                                      \label{a12}
\Sigma(k)=\Delta^{-1}(k)A^{-1}M(k)A\,,
\end{equation}
where $M=(M_{jl})_{j,l=1,2}$.

\medskip

Similarly we can obtain the scattering matrix for an operator
of Neumann type, $H=H^L_{\rm N}$.
We assume for simplicity that the matrix $A$ is diagonal:
$A=(\alpha_j\delta_{jl})_{1\le j,l \le n}$ with real numbers $\alpha_j$, and
that $C$ is a scalar matrix, $C=\gamma I$, $\gamma\in\RR$. In this case we
set
%
%
$$
\Delta_{\rm N}(k):=-|\alpha_1\alpha_2|^2 +(ik-\gamma)\left(|\alpha_2|^2
\tilde Q_{11}(k^2)+|\alpha_1|^2\tilde Q_{22}(k^2)\right)-
(ik-\gamma)^2{\rm det}\,\tilde Q(k^2)\,.
$$
%
Then (\ref{n3.45}) yields
$$
s_{11}(k)=\left[|\alpha_1\alpha_2|^2+
(ik+\gamma)|\alpha_2|^2\tilde Q_{11}(k^2)
-(ik-\gamma)|\alpha_1|^2\tilde Q_{22}(k^2)+\right.
$$
$$
\left.
(k^2+\gamma^2)\det\tilde Q(k^2)\right]
\Delta_{\rm N}^{-1}(k)\,,
$$
$$
s_{22}(k)=\left[|\alpha_1\alpha_2|^2+
(ik+\gamma)|\alpha_1|^2\tilde Q_{22}(k^2)-(ik-\gamma)
|\alpha_2|^2\tilde Q_{11}(k^2)+
\right.
$$
$$
\left.
(k^2+\gamma^2)\det\tilde Q(k^2)\right]
\Delta_{\rm N}^{-1}(k)\,,
$$
$$
s_{12}(k)=2ik \bar \alpha_1\alpha_2\tilde Q_{12}(k^2)
\Delta_{\rm N}^{-1}(k)\,,
$$
\begin{equation}
                           \label{4.9}
s_{21}(k)=2ik \alpha_1\bar\alpha_2\tilde Q_{21}(k^2)
\Delta_{\rm N}^{-1}(k)\,.
\end{equation}

\medskip

\noindent{\bf Remark 8}.  If $-H^0$ is the Laplace--Beltrami operator, then
the scattering matrix (\ref{4.9}) coincides (up to notation) with the one
derived in \cite{ETV}. Moreover, if we put in (\ref{4.9}) $B=C=0$ and
$\alpha_1=\alpha_2=\alpha$, then we get
%
%
$$
s_{21}(k)=\frac{2ik |\alpha|^2 [Q_0(k^2)]_{12}}
{k^2\,{\rm det}\,Q_0(k^2)+
ik |\alpha|^2\left([Q_0(k^2)]_{11}+[Q_0(k^2)]_{22}\right)-|\alpha|^4}\,.
$$
%
This result was obtained by A.~Kiselev \cite{Kis}.

\medskip

Let us list some interesting consequences of (\ref{a8})--(\ref{a11}).
First consider the following permutation of the matrix elements of $A$:
$\alpha_{11}\leftrightarrow \alpha_{12}$, $\alpha_{21}\leftrightarrow
\alpha_{22}$. Then the elements of $\Sigma(k)$ undergo the permutation
$s_{11}\leftrightarrow s_{22}$, $s_{12}\leftrightarrow s_{21}$.
The reason of this effect is intuitively clear: the permutation
$\alpha_{11}\leftrightarrow \alpha_{12}$, $\alpha_{21}\leftrightarrow
\alpha_{22}$ means that we attach the semi-axis
$\RR^+_1$ to the point $q_2$ in place of $q_1$, whereas the semi-axis
$\RR^+_2$ is attached to $q_1$.

Another interesting consequence is related to the conducting properties
of a quantum-mechanical system with the configuration space $\hat X$.
Namely, at zero temperature the ballistic conductance $\sigma(k)$ of an
electric chain consisting of two one-dimensional wires $\Rs^{(1)}$ and
$\Rs^{(2)}$ attached to a mesoscopic device $X$ is given by the
Landauer--B\"uttiker formula
$$
\sigma(k)=\frac{e^2}{\pi\hbar}\frac{T_{12}(k)}{R_1(k)}\,,
$$
%
where $e$ is the electron charge, $\hbar$ is the Planck constant, and
$k^2$ is the Fermi energy \cite{BILP}, \cite{Lan}. For a generic point
$(q_1,q_2)\in X\times X$, $q_1\ne q_2$ and for fixed $z_0\in\sigma(H_0)$,
the function $z\mapsto \det\tQ_0(z)$ has a pole of the second order
at $z_0$. On the other hand, the functions
$z\mapsto \tQ_{jl}(z)$ have poles at most of the first order at the same
point. Therefore, for $T_{12}(k)=T_{21}(k)=|s_{12}(k)|^2$ we have
at a generic point $(q_1,q_2)\in X\times X$, $q_1\ne q_2$,
that $T_{12}(k)=0$ if $k^2\in\sigma(H_0)$. In other words, if
$k^2\in\sigma(H_0)$, then $\sigma(k)=0$. The converse is true, e.g. for
a real operator $H_0$ (i.e. for the operator $H_0$ commuting with
the operator $J$ of complex conjugation: $Jf=\bar f$) at least if the
following conditions are satisfied: (1) the matrix $A$ is diagonal
and $\alpha_{11}\alpha_{22}\ne 0$; (2) ${\rm Im}\,\beta_{12}\ne 0$.
In this case $[Q_0(k^2)]_{12}$ is a real number if
$k^2\notin\sigma(H_0)$, and thus we have the following proposition.

\medskip

{\bf Proposition 8}. {\it Let $\sigma(k)$ be the conductance of an
electric chain consisting of the "wires" $\Rs^{(1)}$ and $\Rs^{(2)}$
attached to the "device" $X$ at some generic points. Suppose that the
Hamiltonian of the device $X$ is a real Schr\"odinger operator $H_0$
of the Dirichlet type. If the conditions $(1)$ and $(2)$ above are
satisfied, then  $\sigma(k)$ vanishes if and only if $k^2$ is an eigenvalue
of $H_0$}. \halm

\medskip

Assume now that ${\rm dim}X\ge2$. If the geodesic distance $r(q_1,q_2)$
between $q_1$ and $q_2$ tends to zero, then at a fixed value of the
energy $k^2$, $k^2\notin \sigma(H_0)$, the numbers $\tQ_{11}(k^2)$ and
$\tQ_{22}(k^2)$ remain bounded, whereas $\tQ_{12}(k^2)$ and
$\tQ_{21}(k^2)$ tend to infinity. Therefore, the conductance $\sigma(k)$
tends to zero (see (\ref{a10}) and (\ref{a11})).
This paradoxical result is intimately related to an unusual
behaviour of the point perturbations of the Schr\"odinger operators
in dimensions 2 or 3. Namely, consider a point perturbation $H_0^B$
of $H_0$ supported on a two-point set $\{q_1,q_2\}$. Then in the sense
of the norm-resolvent convergence, $H_0^B$ tends to the unperturbed
operator $H_0$ as $r(q_1,q_2)\to 0$. Indeed, the above considerations
imply the following assertion: {\it If $z$ is an arbitrary element
of $\rho(H_0)$, then}
\begin{equation}
                           \label{lq}
[Q_0(z)-B]^{-1} \to 0 \quad {\rm as}\,\,\,\,r(q_1,q_2)\to 0\,.
\end{equation}
A discussion of such a property of point perturbations may be found in
\cite{DO}. To overcome the dificulties arising in the limit
$r(q_1,q_2)\to 0$, a renormalization procedure for boundary conditions
has been used \cite{RS}. It is not our intention to discuss here this
subject in detail, we restrict our consideration to some consequences
of (\ref{lq}) for the limiting behaviour of the Schr\"odinger operator
$H$ on $\hat X$.

Applying (\ref{n3.37}) to the matrix $[Q(z)-L]^{-1}$, we get
$$
[Q(z)-L]^{-1}=
\left[
\matrix{ A_{11}&A_{12}\cr
A_{21}&A_{22}\cr}\right],
$$
where
\begin{eqnarray}
A_{11}&=&\left[I-\tilde Q_0^{-1}(z)A\left((-z)^{-1/2}I-C\right)^{-1}
A^*\right]^{-1}\tilde Q_0^{-1}(z)\,,\nonumber\\
A_{12}&=&\tilde Q_0^{-1}(z)A\left[A^*\tilde Q_0^{-1}(z)A-
(-z)^{-1/2}I+C\right]^{-1}\,,\nonumber\\
A_{21}&=&\left[A^*\tilde Q_0^{-1}(z)A-
(-z)^{-1/2}I+C\right]^{-1}A^*\tilde Q_0^{-1}(z)\,,\nonumber\\
A_{22}&=&\left[(-z)^{-1/2}I-C-A^*\tilde Q_0^{-1}(z)A\right]^{-1}\,.
\end{eqnarray}
Now using (\ref{lq}) we show that as $r(q_1,q_2)\to 0$, the operator
$H$ tends in the norm-resolvent sense to the direct sum $H_0\oplus H'$
where $H'$ is a point perturbation (supported in 0) of the free
Hamiltonian $-d^2/dx^2$ on the line $\RR$. It follows from Arnold's Lemma
that in the limit $r(q_1,q_2)\to 0$, we can obtain  any operator of the
form $H_0\oplus H'$ where $H'$
is an {\it arbitrary} point perturbation of $-d^2/dx^2$ supported
on the point 0. In the case of the operator $H=H^L_{\rm D}$, the
limiting scattering matrix can be obtained from (\ref{n3.35}),
it has the form
$$
\Sigma_{\rm D}^{\rm lim}(k)=(ikC-I)(ikC+I)^{-1}\,.
$$
Similarly, if $H=H^L_{\rm N}$, then for the limiting form of the
scattering matrix we obtain from (\ref{n3.44})
$$
\Sigma_{\rm N}^{\rm lim}(k)=(ik+C)(ik-C)^{-1}\,.
$$
We note that in both cases $\Sigma^{\rm lim}(k)$ depends on the block
$C$ of the matrix $L$ only.

In particular, the matrix elements of $\Sigma_{\rm N}^{\rm lim}$ have
the form
\begin{equation}
                                \label{a18}
s_{jl}^{\rm lim}(k)=
\frac{-2ik\gamma_{jl}}{k^2+ik\Tr\,C-\det\,C}\,,\quad j\ne l\,;
\end{equation}

\begin{equation}
                                \label{a19}
s_{jj}^{\rm lim}(k)=
\frac{k^2-ik(\gamma_{jj}-\gamma_{ll})+\det\,C}{k^2+ik\Tr\,C-\det\,C}\,,
\quad j=1,2\,,\quad l\ne j\,.
\end{equation}

Moreover, if $A=I$, then the elements of the scattering matrix
$\Sigma_{\rm N}(k)$ of the initial operator $H^L_{\rm N}$ are the following
$$
s_{11}(k)=
$$
$$
\frac{(k^2-ik(\gamma_{11}-\gamma_{22})+\det\, C)\det\,
\tQ(k^2)+ik(\tQ_{11}(k^2)-\tQ_{22}(k^2))+\Tr\,(C\tQ(k^2))+1}
{\Delta_1(k)}\,,
$$
%
$$
s_{22}(k)=
$$
$$
\frac{(k^2-ik(\gamma_{22}-\gamma_{11})+\det\, C)\det\,
\tQ(k^2)+ik(\tQ_{22}(k^2)-\tQ_{11}(k^2))+\Tr\,(C\tQ(k^2))+1}
{\Delta_1(k)}\,,
$$
%
$$
s_{12}(k)=
\frac{2ik(\tQ_{12}(k^2)-\gamma_{12}\det\, \tQ(k^2))}{\Delta_1(k)}\,,
$$
%
%
$$
s_{21}(k)=
\frac{2ik(\tQ_{21}(k^2)-\gamma_{21}\det\, \tQ(k^2))}{\Delta_1(k)}\,,
$$
%
where
$$
\Delta_1(k):=(k^2+ik\Tr\, C -\det\, C)\det\,
\tQ(k^2)+ik\Tr\,\tQ(k^2)-\Tr\,(C\tQ(k^2))-1\,.
$$
%
It is interesting to compare these elements with those for the
scattering matrix of $H^L_{\rm N}$ in the case of an arbitrary
diagonal matrix $A$ and a scalar matrix $C$ (see (\ref{4.9})).

An important particular case of (\ref{a18}) and (\ref{a19})
arises if we choose the matrix $C$ in the form
$$
C=\left[
\begin{array}{cc}
\gamma& -\gamma\cr
-\gamma& \gamma\cr
\end{array}\right]\,{}_{,}
$$
where $\gamma\in\RR$, $\gamma\ne 0$.
In this case

$$
s_{11}^{\rm lim}(k)=s_{22}^{\rm lim}(k)=
\frac{-ik\gamma^{-1}}{2-ik\gamma^{-1}}\,,
$$
%
%

$$
s_{12}^{\rm lim}(k)=s_{21}^{\rm lim}(k)=
\frac{2}{2-ik\gamma^{-1}}\,.
$$
%
Therefore, the limiting matrix $\Sigma^{\rm lim}(k)$ coincides with the
scattering matrix for the $\delta'$-perturbation
of the free Schr\"odinger operator on the line $\RR$ \cite{AGHH}.
There is a conjecture that the scattering on the $\delta'$-potential
can be realized geometrically \cite{AEL}. Our result shows that
{\it the scattering on the $\delta'$-perturbation can be realized
with an arbitrary accuracy by means of a non-trivial geometric
scattering on an arbitrary compact manifold of dimension} 2 {\it or} 3.

Now we give an example of non-trivial boundary conditions such that the
scattering matrix of the corresponding Schr\"odinger operator $H^{\Lambda}$
in the limit $r(q_1,q_2)\to 0$ (for generic points) has the form

$$
\left[
\begin{array}{cc}
0&1\cr
1&0\cr\end{array}
\right]{}_{,}
$$
i.e. in this limit we obtain a system with zero ballistic resistance
(the condition ${\rm dim}\,X\ge2$ is kept). Namely, let us consider the
boundary conditions of the form (\ref{n1.4a}) where $L$ and $M$ have the
following $2\times 2$-blocks (see (\ref{n3.40a}) for notation):
$X_1=X_2=0$, $Y=I$,
$$
Z=\left[
\begin{array}{cc}
0&0\cr
\zeta&-\zeta\cr
\end{array}
\right]{}_{,}\quad
A_1=\left[
\begin{array}{cc}
\alpha_1&0\cr
\alpha_2&0\cr
\end{array}
\right]{}_{,}\quad
A_2=\left[
\begin{array}{cc}
0&0\cr
\hat\alpha_1&\hat\alpha_2\cr
\end{array}
\right]{}_{,}\quad
C=\left[
\begin{array}{cc}
\gamma&\gamma\cr
0&0\cr
\end{array}
\right]{}_{,}
$$
and $B$ is an arbitrary Hermitian $2\times 2$-matrix. It is easy to prove
that conditions (\ref{n3.40b}) are satisfied iff
$\hat\alpha_j=\zeta\bar\alpha_j$. In this case the scattering matrix
$\Sigma(k)$ is independent of $Z$, $A_2$, and $C$; its elements have the
form:
$$
s_{11}(k)=s_{22}(k)=
$$
\begin{equation}
                         \label{a22}
\frac{|\alpha_1|^2\,\tQ_{11}(k^2)+|\alpha_2|^2\,\tQ_{22}(k^2)
-\bar\alpha_1\alpha_2\tQ_{12}(k^2)-\bar\alpha_2\alpha_1\tQ_{21}(k^2)}
{|\alpha_1|^2\,\tQ_{11}(k^2)+|\alpha_2|^2\,\tQ_{22}(k^2)
-\bar\alpha_1\alpha_2\tQ_{12}(k^2)-\bar\alpha_2\alpha_1
\tQ_{21}(k^2)-2ik^{-1}\det\,\tQ(k^2)}\,,
\end{equation}
$$
s_{12}(k)=s_{21}(k)=
$$
\begin{equation}
                         \label{a23}
\frac{2ik^{-1}\det\,\tQ(k^2)}
{2ik^{-1}\det\,\tQ(k^2)- |\alpha_1|^2\,\tQ_{11}(k^2)-|\alpha_2|^2\,
\tQ_{22}(k^2)
+\bar\alpha_1\alpha_2\tQ_{12}(k^2)+\bar\alpha_2\alpha_1\tQ_{21}(k^2)}\,.
\end{equation}

It is curious that the conductance of a system with the Hamiltonian
$H^{\Lambda}$ is in some sense reciprocal to the one described in
Proposition~8. In fact, (\ref{a22}) and (\ref{a23}) show immediately
that the following proposition is true.

\medskip

{\bf Proposition 9}. {\it Suppose that the semi-axes $\Rs^{(1)}$ and
$\Rs^{(2)}$ are attached to $X$ in generic points and that the Schr\"odinger
operator $H^{\Lambda}$ on $\hat X$ is given as above. Then $\sigma(k)=\infty$
$($i.e. the system $\hat X$ is a superconductor at the energy level $k^2)$
if and only if $k^2\in\sigma(H_0)$. Moreover, $\sigma(k)=0$ if and only if
$k^2$ is an energy level for the point perturbation $H_0^B$ of $H_0$}. 


\vskip 5ex

\noindent{\bf 5.~A few examples}

\bigskip

\noindent Here some examples of Schr\"odinger operators $H_0$ on a compact
manifold $X$ of constant curvature are collected for which we can give an
explicit form of the $\Q$-matrix $Q_0$ and, hence, get an explicit
expression
for the scattering matrix $\Sigma(k)$ via (\ref{n3.35}) or (\ref{n3.44}).
Recall that for $j\ne l$
\begin{equation}
                   \label{n4.0}
[Q_0(z)]_{jl}=G_0(q_j,q_l;z)\,,
\end{equation}
where $G_0(x,y;z)$ is the Green function of $H_0$.
Therefore, as a rule, only the diagonal terms $Q_0(z)]_{jj}$ are written
explicitly below.

\medskip

\noindent (1) {\it Ring $\SS_a$}

\medskip

Let $X$ be a ring $\SS_a$ (i.e. a circle) of radius $a$. It is easy to show
that the Green function for the Schr\"odinger operator of a free
charged particle
%
 %
$$                         \label{n4.1}
H_0=-\frac{1}{a^2}\frac{d^2}{d\phi^2}
$$
%
($\phi\in[0\,,2\pi)$ being the polar coordinate on $\SS_a$) has the form
%
%
$$
G_0(\phi,\phi';z)=-\frac{1}{2\sqrt{z}}\frac{\cos a\sqrt{z}(\phi'-\phi\pm \pi)}
{\sin\pi a\sqrt{z}}\,,
$$
%
where the sign "plus" is taken if $\phi\ge \phi'$, otherwise we take "minus". The diagonal elements
of the matrix $Q_0$ have the form:
\begin{equation}
                        \label{n4.3}
[Q_0(z)]_{jj}=G_0(q_j,q_j;z)\,.
\end{equation}

\medskip

\noindent (2) {\it Aharonov--Bohm ring}

\medskip

Consider a ring $\SS_a$ of radius $a$ located in an axially symmetric
magnetic field perpendicular to the plane of the ring. Let $\Phi$ be
the total magnetic flux through the ring. Put $\vartheta=\Phi/\Phi_0$ where
$\Phi_0$ is the quantum of the magnetic flux: $\Phi_0=2\pi\hbar c/|e|$.
Then the Schr\"odinger operator for a charged particle in the system
considered has the form
%
%
$$
H_0=\frac{1}{a^2}\left(-i\frac{d}{d\phi}+\vartheta\right)^2.
$$
For the Green function we have \cite{GPa}:
$$
G_0(\phi,\phi';z)=
%
%
\frac{1}{4\sqrt{z}}\left[
\frac{\exp\left(i(\phi'-\phi\pm\pi)(\vartheta-a\sqrt{z})\right)}
{\sin\pi(\vartheta-a\sqrt{z})}-
\frac{\exp\left(i(\phi'-\phi\pm\pi)(\vartheta+a\sqrt{z})\right)}
{\sin\pi(\vartheta+a\sqrt{z})}
\right]
$$
\medskip
(the choice of the signs is as in the previous example). In the 
considered case the matrix $Q_0$ is given by 
(\ref{n4.0}) and (\ref{n4.3}) again.

\medskip

\noindent (3) {\it Flat torus $\TT^d$ $(d=2$ or $3)$}

\medskip

Let $\Lambda_d$ be a lattice in $\RR^d$ with generators $\va_1$,..., $\va_d$:
%
%
$$
\Lambda_d=\{n_1\va_1+\ldots+n_d\va_d:\,n_j\in\ZZ,\,j=1,\ldots,d\}\,,
$$
%
and let $\Gamma_d$ be the dual lattice for $\Lambda_d$, i.e. $\Gamma_d$
be the lattice with generators $\vb_1,\ldots,\vb_d$ obeying the condition
$\va_j\vb_k=2\pi\delta_{jk}$.
Denote by $F_d$ the elementary cell for $\Lambda_d$:
%
%
$$
\displaystyle
F_d=\left\{x_1\va_1+\ldots+x_d\va_d:\,-\frac{1}{2}\le
x_j<\frac{1}{2}\right\}\,
$$
%
and fix points $q_1,\ldots,q_n$ from $F_d$. Let $H_0=-\Delta_X$ where
$X$ is the the torus $\TT^d=\RR^d/\Lambda_d$. Choosing points
$q_1,\ldots,q_n\in \TT^d$, we have \cite{AGHH}:
\begin{equation}
                         \label{n4.8}
[Q_0(z)]_{jl}=
\displaystyle
\cases{
\displaystyle
v_d^{-1}\lim\limits_{\omega\to\infty}\sum\limits_{\gamma\in\Gamma_d,
\,|\gamma|\le\omega}
\frac{e^{i\gamma(q_j-q_l)}}{|\gamma|^2-z}\,, & if $j\ne l$;\cr
\noalign{\medskip}
\displaystyle
(2\pi)^{-d}\lim\limits_{\omega\to\infty}\left[
\sum\limits_{\gamma\in\Gamma_d,\,|\gamma|\le\omega}
\frac{\hat v_d}{|\gamma|^2-z}-\xi_d(\omega)\right]\,, & if $j=l$\,.\cr}
\end{equation}

\noindent Here $v_d$ and $\hat v_d$ are the volumes of the tori
$\RR^d/\Lambda^d$ and $\RR^d/\Gamma^d$ respectively; the functions
$\xi_d$ $(d=2,3)$ have the form

%
%
$$
\xi_d(\omega)=\cases{2\pi\ln\omega\,, & if $d=2$;\cr
\noalign{\medskip}
4\pi\omega\,, &if $d=3$\,.\cr}
$$
%

Using either the eigenfunction expansion for the Laplace operator on $\TT^d$
or the Poisson summation formula we can get an convergent absolutely  series
expansion for $Q_0(k^2)_{jl}$ (see \cite{Kar} for the case $d=3$):

\begin{equation}
                    \label{n4.10}
[Q_0(k^2)]_{jl}=
\displaystyle
(1+z)v_d^{-1}\sum\limits_{\gamma\in\Gamma_d}\frac{e^{i\gamma(q_j-q_l)}}
{(|\gamma|^2-z)(|\gamma|^2+1)}+\kappa_d(q_j-q_l)\,.
\end{equation}

\noindent Here the functions $\kappa_d$ are defined as follows:
If $d=2$, then

%
%
$$
\kappa_2(x)=
\displaystyle
\cases{
\displaystyle
\frac{1}{2\pi}\sum\limits_{\lambda\in\Lambda_d}K_0(|x+\lambda|)
\,, & if $x\notin \Lambda_d$;\cr
\noalign{\medskip}
\displaystyle
\frac{1}{2\pi}\left[
\sum\limits_{\lambda\in\Lambda_d,\,\lambda\ne 0}K_0(|\lambda|)+\ln2-C_E
\right]\,, & if $x\in\Lambda_d$\,,\cr}
$$
%

\noindent where $K_0$ is the Macdonald function (i.e. the modified
Bessel function of the third kind) and $C_E$ is the Euler constant.
In the case $d=3$ we have

%
%
$$
\kappa_3(x)=
\displaystyle
\cases{
\displaystyle
\frac{1}{4\pi}\sum\limits_{\lambda\in\Lambda_d}
\frac{e^{-|x+\lambda|}}{|x+\lambda|}\,, & if $x\notin \Lambda_d$;\cr
\noalign{\medskip}
\displaystyle
\frac{1}{4\pi}\left[
\sum\limits_{\lambda\in\Lambda_d,\,\lambda\ne 0}
\frac{e^{-|\lambda|}}{|\lambda|}-1\right]\,, & if $x\in\Lambda_d$\,.\cr}
$$
%

\medskip

\noindent (4) {\it Flat torus with Aharonov--Bohm fluxes}

\medskip

Consider the torus $\TT^d$ as the product of $d$ Aharonov--Bohm rings
$\SS_{a_j}$ with fluxes $\vartheta_j$ ($j=1,\ldots,\,d$). Let
$$
H_j=\frac{1}{a_j^2}\left(-i\frac{d}{d\phi}+\vartheta_j\right)^2\,,
$$
and

%
%
$$
H_0=\cases{
H_1\otimes I_2+I_1\otimes H_2\,,& if $d=2$\,;\cr
\noalign{\medskip}
H_1\otimes I_2\otimes I _3+I_1\otimes H_2\otimes I_3+
I_1\otimes I_2\otimes H_3\,,& if $d=3$\,.\cr}
$$
%

\medskip

\noindent The operator $H_0$ may be considered as the Schr\"odinger operator
on a torus $\TT^d$ with a non-uniform magnetic field.
Denote by $\vartheta$ the vector $(\vartheta_1,\ldots,\vartheta_d)$, then
the $\cal Q$-function  $Q_0$ now takes the form

%
%
$$
[Q_0(z)]_{jl}=
\displaystyle
\cases{
\displaystyle
v_d^{-1}\lim\limits_{\omega\to\infty}\sum\limits_{\gamma\in\Gamma_d,\,
|\gamma+\theta|\le\omega}
\frac{e^{i(\gamma+\vartheta)(q_j-q_l)}}
{|\gamma+\vartheta|^2-z}\,, & if $j\ne l$;\cr
\noalign{\medskip}
\displaystyle
(2\pi)^{-d}\lim\limits_{\omega\to\infty}\left[
\sum\limits_{\gamma\in\Gamma_d,\,|\gamma+\theta|\le\omega}
\frac{\hat v_d}{|\gamma+\vartheta|^2-z}-
\xi_d(\omega)\right]\,, & if $j=l$\,,\cr}
$$
%

\noindent or

%
%
$$
[Q_0(z)]_{jl}=
\displaystyle
(1+z)v_d^{-1}\sum\limits_{\gamma\in\Gamma_d}
\frac{e^{i(\gamma+\vartheta)(q_j-q_l)}}
{(|\gamma+\vartheta|^2-z)(|\gamma+\vartheta|^2+1)}+
\kappa_{d,\vartheta}(q_j-q_l)\,.
$$
%

\noindent Now the functions $\kappa_{d,\vartheta}$ ($d=2$, $3$) are defined as follows:

%
%
$$
\kappa_{2,\vartheta}(x)=
\displaystyle
\cases{
\displaystyle
\frac{1}{2\pi}\sum\limits_{\lambda\in\Lambda_d}K_0(|x+\lambda|)
e^{-i\vartheta\lambda}
\,, & if $x\notin \Lambda_d$;\cr
\noalign{\medskip}
\displaystyle
\frac{1}{2\pi}\left[
\sum\limits_{\lambda\in\Lambda_d,\,\lambda\ne 0}K_0(|\lambda|)
e^{-i\vartheta\lambda}+\ln2-C_E
\right]\,, & if $x\in\Lambda_d$\,;\cr}
$$
%

%
%
$$
\kappa_{3,\vartheta}(x)=
\displaystyle
\cases{
\displaystyle
\frac{1}{4\pi}\sum\limits_{\lambda\in\Lambda_d}
\frac{e^{-|x+\lambda|-i\vartheta\lambda}}{|x+\lambda|}
\,, & if $x\notin \Lambda_d$;\cr
\noalign{\medskip}
\displaystyle
\frac{1}{4\pi}\left[
\sum\limits_{\lambda\in\Lambda_d,\,\lambda\ne 0}
\frac{e^{-|\lambda|-i\vartheta\lambda}}{|\lambda|}-1
\right]\,, & if $x\in\Lambda_d$\,.\cr}
$$
%

\medskip

\noindent (5) {\it Flat torus $\TT^2$ with a perpendicular uniform
magnetic field}

\medskip

Consider the Euclidean plane $\RR^2$ with the lattice $\Lambda_2$ and let
${\bf B}$ be a uniform magnetic field that is perpendicular to the plane
and has the strength $B$. Denote by $\vartheta$ the number of the magnetic
flux quanta through the elementary cell $F_2$: $\vartheta=Bv_d/\Phi_0$.
The Green function $G^0$ for the Schr\"odinger operator of a charged
particle on the plane $\RR^2$ with the field ${\bf B}$ has the form:
$$
\displaystyle
G^0(x,y;z)=\frac{1}{4\pi}\Gamma\left(\frac{1}{2}-\frac{v_d z}
{4\pi |\vartheta|}\right)\times
$$
%
$$
\displaystyle
\exp\left[-i\pi \vartheta v_d^{-1}x\wedge y-
\frac{\pi|\vartheta|}{2v_d}(x-y)^2\right]
\Psi\left(\frac{1}{2}-\frac{v_dz}{4\pi|\vartheta|}\,,
 1\,; \frac{\pi|\vartheta|}{v_d}\,\,(x-y)^2\right)\,,
$$

\medskip

\noindent where $\Gamma(z)$ is the Euler $\Gamma$-function, $\Psi(a,c;z)$
is the Tricomi function (the confluent hypergeometric function), and
$x\wedge y=x_1y_2-x_2y_1$ is the standard symplectic product in $\RR^2$.
Let the following quantization condition be satisfied:
{\it the number $\vartheta=Bv_d/\Phi_0$ of the flux quanta through the
cell $F_2$ is an integer}. Then we can consider the corresponding
magnetic Schr\"odinger operator on the torus $\TT^2$. Using results
from \cite{Gey} we obtain for the Krein ${\cal Q}$-matrix:
\begin{equation}
                            \label{n4.19}
[Q_0(z)]_{jl}=\sum\limits_{\lambda\in\Lambda_2,\,\lambda\ne 0}
G^0(\lambda+q_j,q_l;z)\exp\left[\pi i\vartheta v_d^{-1}(q_j\wedge \lambda)-
\pi i \vartheta\lambda_1\lambda_2\right] +\xi_{jl}(z)\,.
\end{equation}
Here
%
%
$$
\xi_{jl}(z)=\cases{G^0(q_j,q_l;z)\,, &if $j\ne l$;\cr
\noalign{\medskip}
\displaystyle
-\frac{1}{4\pi}\left[\psi\left(\frac{1}{2}-\frac{v_dz}
{4\pi|\vartheta|}\right)+
\ln\left(\pi|\vartheta|v_d^{-1}\right)+2C_E\right]
\,, &if $j=l$\,,\cr}
$$
%
where $\psi(z)$ is the digamma function (the logarithmic derivative of the
$\Gamma$-function). Note that in (\ref{n4.19}), $\lambda_1$, $\lambda_2$
are the coordinates of $\lambda$ in the basis $\va_1$, $\va_2$ of
$\Lambda_2$: $\lambda=\lambda_1\va_1+\lambda_2\va_2$.

\medskip

\noindent (6) {\it Sphere $\SS^2_a$}

\medskip

Let $X$ be a two-dimensional sphere $\SS_a^2$ of radius $a$, then
the Green function for the Schr\"odinger operator $H_0$ of a free particle on $X$,
$H_0=-\Delta_X$, has the form \cite{GS}:
%
$$
\displaystyle
G_0(x,y;z)=
-\frac{1}{4\cos\left(\pi t(z)\right)}\,\,
{\cal P}_{-\frac{1}{2}+t(z)}\left(-\cos\frac{r(x,y)}{a}\right)
\,,
$$
%

\noindent where ${\cal P}_a(z)$ is the Legendre function and
$$
t(z)=\frac{1}{2}\sqrt{1+4a^2z}\,.
$$
Therefore, for every $j$
$$
[Q_0(z)]_{jj}=-\frac{1}{4\pi}\left[\psi\left({1\over2}+t(z)\right)+
\psi\left({1\over2}-t(z)\right)-2\ln(2a)+2C_E\right]=
$$
%
$$
-\frac{1}{2\pi}\left[\psi\left({1\over2}+t(z)\right)-
{\pi\over2}{\rm tg}\,(\pi t(z))-\ln(2a)+C_E\right]\,.
$$
%

\medskip

\noindent (7) {\it Sphere $\SS^3_a$}

\medskip

Consider now a three-dimensional sphere $X=\SS^3_a$ of radius $a$. Then
the Green function for the Schr\"odinger operator $H_0$ of a free particle
on $X$, $H_0=-\Delta_X$, reads \cite{GS}:

$$
\displaystyle
G_0(x,y;z)=
%
%
\frac{1}{4\pi a\sin\frac{r(x,y)}{a}}
\left[\cos\frac{r(x,y)\sqrt{a^2z+1}}{a}-
\sin\frac{r(x,y)\sqrt{a^2z+1}}{a}
{\rm ctg}\pi\sqrt{a^2z+1}\right]\,.
$$
%

\noindent Therefore, for every $j$
%
%
$$
\displaystyle
[Q_0(z)]_{jj}=
-\frac{\sqrt{a^2z+1}}{4\pi a}\,\,{\rm ctg}\pi \sqrt{a^2z+1}\,.
$$
%

\medskip

\noindent (8) {\it Compact manifold of constant negative curvature}

\medskip

Let now $X$ be a compact $d$-dimensional manifold of constant negative
curvature (with sectional curvature $-a^{-2}$ for some $a>0$). We shall
consider $X$ as a quotient $\HP^d/\Gamma$, where $\HP^d$ is the
$d$-dimensional Lobachevsky space (i.e. the complete simply connected
$d$-dimensional Riemannian manifold of constant negative curvature) and
$\Gamma$ is a cocompact discontinuous group of motions in $\HP^d$.
Denote by $G^0_d$ the Green function for the Laplace--Beltrami operator
on $\HP^d$. Recall that
$$
G^0_d (x,y;z)=
$$
%
$$
\displaystyle
\cases{\displaystyle
\frac{\Gamma^2(s_2(z))}{4\pi\Gamma(2s_2(z))}\left[{\rm cosh}\,
\frac{r(x,y)}{2a}\right]^{-2s_2(z)}
F\left(s_2(z),s_2(z);2s_2(z);{\rm cosh}^{-2}\,
\frac{r(x,y)}{2a}\right)\,,& if $d=2$;\cr
\noalign{\bigskip}\displaystyle
\frac{\exp\left[a^{-1}r(x,y)(1-s_3(z))\right]}{4\pi a \,{\rm sinh}\,
\left(a^{-1}r(x,y)\right)}\,, &if $d=3$
\cr}
$$
%

\medskip

\noindent (see \cite{Els}, \cite{EGM}). Here $F(a,b;c;z)$ is the
Gauss hypergeometric function and
\begin{equation}
                 \label{4.25}
\displaystyle
s_d(z)=\frac{d-1+\sqrt{(d-1)^2-4a^2z}}{2}, \quad d=2\,,3\,.
\end{equation}
Let $H_0$ be a Schr\"odinger operator on $X$ of the form $H_0=-\Delta_X$.
If ${\rm Re}\,s_d(z)$ is sufficiently large, then there is an expansion of
the Green function $G_0(x,y;z)$ for $x\ne y$ into an absolutely convergent
series \cite{Fad}, \cite{Els}, \cite{EGM}:
\begin{equation}
                 \label{n4.26}
G_0(x,y;z)=
\sum\limits_{\gamma\in\Gamma}G_d^0(x,\gamma y;z)\,.
\end{equation}

\medskip

\noindent To find $G_0(x,y;z)$ for an arbitrary
$z\in \CC\setminus \sigma(H_0)$ we choose a number $z'={\rm Re}\,z+ik$,
where $k\in\RR$ is so large that the series (\ref{n4.26}) absolutely
converges at $z=z'$.
%
%
%
%
%
%
%
%
%
%
Then the Neumann series
%
%
$$
R_0(z)=\sum\limits_{n=0}^{\infty}(z-z')^nR_0^{n+1}(z')\,
$$
%
gives the desired value $R_0(z)$ and $G_0(x,y;z)$ may be found as an
infinite sum of iterated integral kernels $G_0(x,y;z')$.

To find the Krein ${\cal Q}$-function we use (\ref{n4.26}) again.
If ${\Re}\,s_d(z)$ is sufficiently large, then
\begin{equation}
                               \label{n4.30}
[Q_0(z)]_{jj}=\sum\limits_{\gamma\in\Gamma,\,\gamma\ne 1}
G_d^0(q_j,\gamma q_j;z)+\kappa_d(z)\,,
\end{equation}
where
%
%
$$
\kappa_d(z)=\cases{
\displaystyle
-\frac{1}{2\pi}\left[\psi(s_2(z))-\ln 2a+C_E\right]\,, &if $d=2$;\cr
\noalign{\medskip}
\displaystyle
-\frac{1}{4\pi a}\sqrt{1-a^2z}\,,& if $d=3$\,.\cr}
$$
%
\smallskip
\noindent To find $[Q_0(z)]_{jj}$ at an arbitrary point $z$,
$z\notin \sigma(H_0)$, we fix $z_0\in\RR$, $z_0<0$ such that
$[Q_0(z_0)]_{jj}$ is given by (\ref{n4.30}). Using the Hilbert resolvent
identity and taking into consideration that the integral kernel for
$R_0(z)R_0(z_0)$ is continuous \cite{Els}, \cite{EGM}, we get:
%
%
$$
[Q_0(z)]_{jj}=[Q_0(z_0)]_{jj}+
(z-z_0)\int\limits_X G^0_d(q_j,x;z)G^0_d(x,q_j;z_0)\,d\lambda(x)\,.
$$

\medskip

\noindent{\bf Remark 9}. In some sense (\ref{n4.30}) is an analogue of
(\ref{n4.8}) for the space of constant negative curvature. Let us consider
for simplicity the case of one horn ($n=1$, $q_1=q$) and try to transform
(\ref{n4.30}) to an equality similar to (\ref{n4.10}) hoping to get a more
convenient expression. First note that in general $Q_0(z)$ depends on $q$:
$Q_0(z)=Q_0(z,q)$. But the Poisson summation formula gives us an averaged
value $Q_0^{\rm av}(z)$ of $Q_0(z,q)$:
$$
Q_0^{\rm av}(z):=({\rm vol}\,X)^{-1}\int\limits_X Q_0(z,q)\,dq\,.
$$
If $X$ is a homogeneous manifold, then $Q_0^{\rm av}(z)$ is independent
of $q$ and $Q_0(z)=Q_0^{\rm av}(z)$. Therefore, in the case of the torus
$\TT^d$, $Q_0(z)$ is given by (\ref{n4.10}). Let now $X$ be a compact surface
of constant negative curvature. In this case the role of the Poisson
summation formula is played by the Selberg trace formula. Using the Selberg
formula in the form obtained by P.~Cartier and A.~Voros  \cite{CV}  we get
an explicit expression for $Q_0^{\rm av}(z)$ up to an additive constant~$c$:
$$
Q_0^{\rm av}(z)=(2-2g) \psi\left(s_2(z)\right)+
\frac{1}{\sqrt{1-4a^2z}}
\frac{{\cal Z}'_X\left(s_2(z)\right)}
{{\cal Z}_X\left(s_2(z)\right)}+c\,,
$$
where $g$ is the genus of $X$, ${\cal Z}_X(s)$ is the Selberg zeta function
for $X$ \cite{Hej}, \cite{Sel}, and $s_2(z)$ is given by (\ref{4.25}).
Note that without loss of generality we can put $c=0$, otherwise we add
$c$ to the parameter $\beta$ in (\ref{4.2}) and (\ref{4.3}).

\bigskip

\noindent (9) {\it Compact Riemann surface of constant negative curvature
with a uniform magnetic field}

\medskip

Consider the Lobachevsky plane $\HP^2$ with a uniform magnetic field
${\bf B}$ of strength $B$ perpendicular to the plane \cite{Com}. Using
the Poincar\'e half-plane realization for $\HP^2$
($\HP^2=\{x\in\RR^2:\,x_2>0\}$ with the metric $r(x,y)=
a\,{\rm cosh}^{-1}(1+(2x_2y_2)^{-1}|x-y|^2)$), we have the following
representation for the Green function $G^0(x,y;z)$ of the magnetic
Schr\"odinger operator on $\HP^2$ \cite{Els}, \cite{Com}:
$$
\displaystyle
G^0(x,y;z)=
\frac{\exp(ib\phi)}{4\pi}\frac{\Gamma(t(z)+b)\Gamma(t(z)-b)}
{\Gamma(2t(z))}\times
$$
%
$$
\displaystyle
\left[{\cosh}\,\frac{r(x,y)}{2a}\right]^{-2t(z)}
F\left(t(z)+b\,,\,t(z)-b\,;\,2t(z)\,;\,{\rm cosh}^{-2}\,
\frac{r(x,y)}{2a}\right)\,,
$$
%
where
$$
\varphi=2{\rm arctg}\frac{x_1-y_1}{x_2+y_2}\,,
$$
$$
t(z)=\frac{1}{2}\left(1+\sqrt{1-4(a^2z-b^2)}\,\right)\,,
$$
$$
b=Ba^2/\Phi_0\,.
$$

Let $S_\Gamma$ be the area of a fundamental domain for $\Gamma$ and
suppose that $BS_{\Gamma}/\Phi_0$ is an integer. Then one can define
the magnetic Schr\"odinger operator $H_0$ on the manifold
$X=\HP^2/\Gamma$, and its Green function has the form (\ref{n4.26})
for sufficiently large ${\rm Re}\,t(z)$ \cite{Els}. For this $t(z)$ we
obtain, using a result from \cite{BG1},
%
%
$$
[Q_0(z)]_{jl}=\sum\limits_{\gamma\in\Gamma,\,\gamma\ne 1}
G^0(q_j,\gamma q_l;z)+\xi_{jl}(z)\,,
$$
%
where
%
%
$$
\xi_{jl}(z)=\cases{G^0(q_j,q_l;z)\,, &if $j\ne l$;\cr
\noalign{\medskip}
\displaystyle
-\frac{1}{4\pi}\left[\psi(t(z)+b)-\psi(t(z)-b)-2\ln 2a+2C_E\right]\,,
&if $j=l$\,.\cr}
$$

To define the ${\cal Q}$-matrix at other points of $\CC\setminus\sigma(H_0)$
it is sufficient to apply the method presented in Example 7.

\bigskip

\noindent{\bf Acknowledgments}

\noindent
The authors are grateful to P.~Exner for useful discussions and for the
possibility to get acquainted with the results of \cite{ETV} before
publication.

We gratefully acknowledge
grants of DFG (No. 436 RUS 113/572/1) and INTAS (No. 00-257). 
The second named author
also very grateful to RFBR (Grant No. 02-01-00804) and to 
the SFB 288 for a financial support, and
to  Humboldt University of Berlin for warm hospitality
during the preparation of this paper.

\vskip 1truecm

\baselineskip=12pt


\end{document}